\documentclass[aps,prb,twocolumn,groupedaddress,superscriptaddress,showpacs,floatfix,longbibliography]{revtex4-2}
\usepackage{graphicx}
\usepackage{physics}
\usepackage{float}
\usepackage{xcolor}
\usepackage{amsfonts}
\usepackage{amsmath}
\usepackage{amssymb}
\usepackage{rotating}
\usepackage{bm}
\usepackage{bbm}
\usepackage{kantlipsum}
\usepackage{braket}
\usepackage{dsfont}
\usepackage{xtab,afterpage}
\usepackage{lipsum}
\usepackage{comment}

\usepackage{placeins}

\usepackage{makecell}

\def\be{\begin{eqnarray}}
\def\ee{\end{eqnarray}}

\usepackage[unicode=true, bookmarks=false,breaklinks=false,pdfborder={0 0 1},backref=false,colorlinks=false]{hyperref}
\hypersetup{hypertex}   
\usepackage{breakurl}
\hypersetup{backref,colorlinks=true,citecolor=blue,linkcolor=blue,filecolor=blue,urlcolor=blue}
\usepackage{bibunits}
\defaultbibliography{ref}
\defaultbibliographystyle{apsrev4-2}

\begin{document}

\title{Identifying strong correlation using only the Kohn-Sham density of one-electron states}

\author{Daniel D. Rivera}
\affiliation{Department of Physics and Engineering Physics, Tulane University, New Orleans, LA 70118, USA}
\affiliation{Instituto de Física, Universidade de São Paulo, 05315-970 São Paulo, São Paulo, Brasil}

\author{Gustavo M. Dalpian}
\affiliation{Instituto de Física, Universidade de São Paulo,  05315-970 São Paulo, São Paulo, Brasil}

\author{John P. Perdew}
\email{perdew@tulane.edu}
\affiliation{Department of Physics and Engineering Physics, Tulane University, New Orleans, LA 70118, USA}

\begin{abstract}
Strongly correlated systems have long been a central and highly non-trivial topic in condensed matter physics. At the non-interacting level, strong correlation can be associated with powerful (near) degeneracies between occupied and unoccupied states, which leads to a high density of states near the Fermi level in metallic configurations. Such regimes are commonly treated with beyond-density functional theory (DFT) approaches, such as DFT+U or DFT+DMFT while maintaining symmetric configurations. Here, we explore the hypothesis that symmetry breaking in the Kohn–Sham (KS) non-interacting system can qualitatively account for the energetic effects of strong correlation in the corresponding interacting system within standard DFT. By lifting near-degeneracies around the Fermi level, symmetry breaking diminishes the potential correlation effects, reducing the need for an explicit treatment of electron correlation, transforming an otherwise strongly correlated symmetric configuration into a normally correlated one, thus avoiding the need for interacting methods beyond DFT. This naturally connects nonmagnetic to magnetic states. We apply this idea to both strongly and normally correlated metals and observe that spin symmetry breaking leads to a pronounced reduction of the density of states at the Fermi level and a significant lowering of the total energy in strongly correlated cases. To describe the degree of correlation that the interacting system would have relative to the KS state, we introduce a correlation parameter ($\Gamma$), defined as the ratio between the Kohn-Sham density of one-electron states at the Fermi level and that of a corresponding uniform electron gas. This parameter distinguishes strongly correlated systems, which would require explicit treatment, from normally correlated ones, which do not.
\end{abstract}

\pacs{}
\maketitle

\begin{bibunit}


Strong correlation has long been recognized as difficult to capture within standard condensed matter theoretical frameworks, including Density Functional Theory (DFT) and Coupled Cluster Theory \cite{martin2004electronic,perdew-zunger-2025,Kaplan2023,scuseria2015}. More recently, symmetry breaking has emerged as a promising alternative route for describing strongly correlated systems \cite{Perdew2021,Zunger2022,jianweisun2018,Zunger2025,Wang2021,Zhang2020,Perdew2022,perdew-zunger-2025,Perdew2025,Malyi2023,Joshi2024,Jia2025,ZiKuiLiu2024,Maniar2024,wang2020}. By allowing symmetry breaking in different degrees of freedom—such as spin, atomic positions, and electric dipoles \cite{Zunger2022}—it becomes possible to reproduce key correlation-driven phenomena, including Mott insulating phases \cite{Zunger2025,Zhang2020}, band-gap openings \cite{jianweisun2018,Zunger2022,perdew-zunger-2025}, and quasiparticle mass renormalization \cite{Wang2021}. Remarkably, these effects can be captured without explicitly invoking on-site electron–electron repulsion terms, which were traditionally considered essential for describing strong correlation. This naturally raises a fundamental question: if symmetry breaking successfully captures these phenomena, where is strong correlation effectively encoded once symmetry breaking is taken into account \cite{Zunger2022}?

\begin{figure}[t!]
    \centering
    \includegraphics[width=1.05\linewidth]{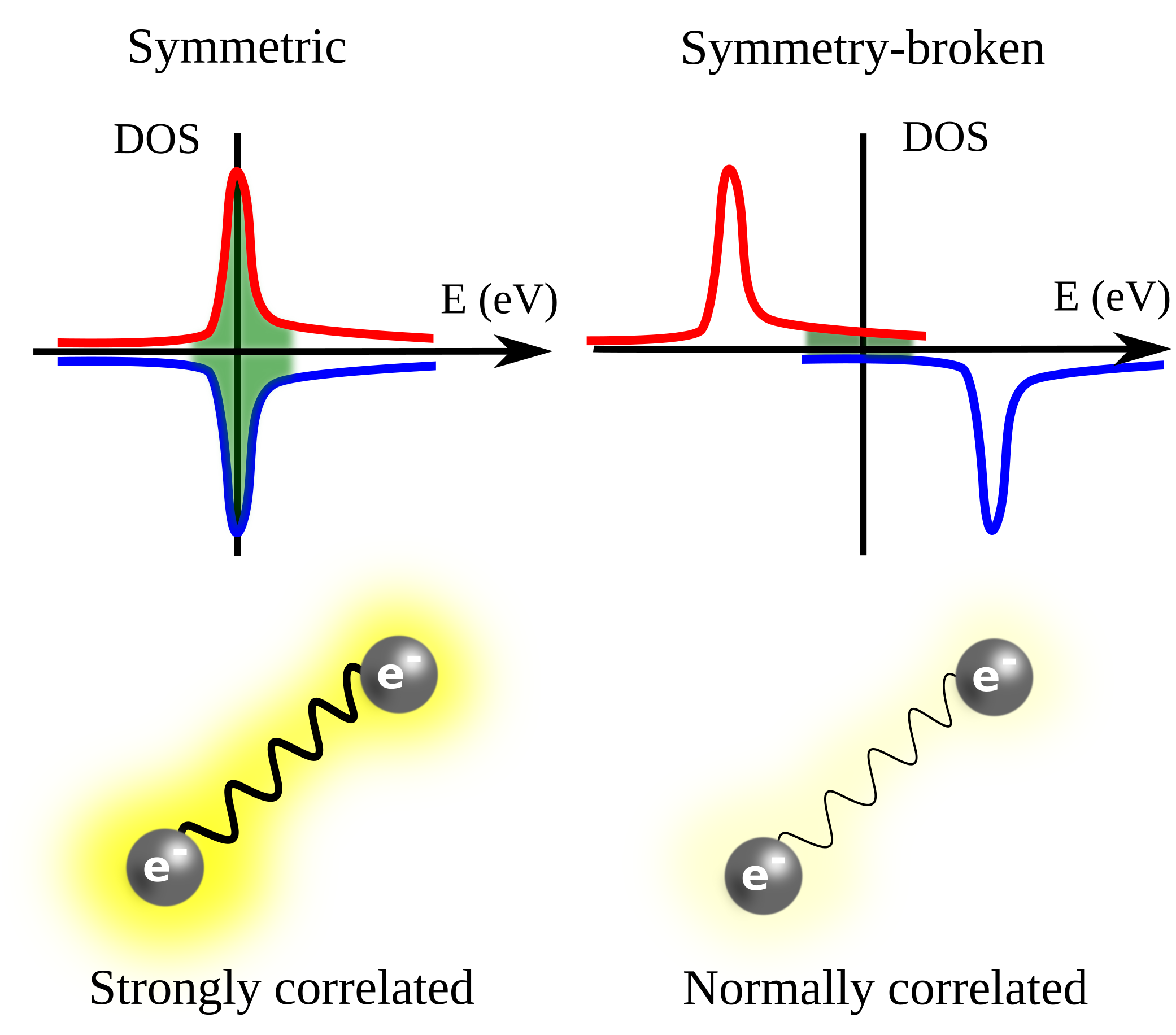}
    \caption{Spin symmetry breaking effect on the density of states. The green shaded region marks the vicinity of the Fermi level. Positive/red (negative/blue) density of states correspond to spin up (down). The figure illustrates that symmetric states exhibit a larger number of degenerate electronic states near the Fermi level. Symmetry breaking reduces the number of quasi-degenerate states in this region, thereby reducing the effective impact of the missing electron–electron interactions at the non-interacting level.}
    \label{fig:figure0}
\end{figure}

At the non-interacting level, strong correlation is associated with (near) degeneracies among relevant Slater determinants in a configuration interaction framework, or, equivalently, with a small energy separation between occupied and unoccupied one electron states in the leading determinant of a single-reference description, typically of \textit{d} or \textit{f} character \cite{Perdew2025, perdew-zunger-2025}. In metals, this could manifest as a high density of states near the Fermi level for both occupied and unoccupied states \cite{perdew-zunger-2025,Perdew2025}. Once the electron–electron interaction is turned on, as a perturbation for instance, a smaller energy separation between the unperturbed occupied and unoccupied single-particle states implies a stronger impact of the perturbation, in this case the interaction itself, ultimately leading to a strongly correlated solid \cite{mqm}. This is illustrated in figure \ref{fig:figure0}. Symmetry breaking results in the lifting of degeneracies, including accidental ones, thereby increasing the energy separation between relevant electronic states, potentialy decreasing the need for the explicit inclusion of electron-electron interaction terms in the non-interacting hamiltonian. 

In density functional theory (DFT), an auxiliary Kohn–Sham (non-interacting) system is introduced to reproduce the ground-state electron density of the interacting electronic system \cite{kohn1964,Kohn1965}. In principle, if the exact exchange–correlation functional was known, this construction would yield both the exact ground-state density and the exact total energy of the interacting system. In practice, however, the exchange–correlation functional must be approximated in a way that accounts only for normal correlation effects \cite{Kaplan2023,Perdew1996,Furness2020}. This limitation arises because density functional approximations are typically constructed using normally correlated systems as reference norms \cite{Kaplan2023}, which leads to functionals that lack the ingredients needed to properly capture strongly correlated physics. As a result, the total energy may fail to capture the stabilization due to (very negative) strong correlation, leading to energies that are less negative than they should be in strongly correlated regimes, making the theory less reliable when strong correlation effects are dominant \cite{martin2004electronic,Kaplan2023}. To address this issue, corrective approaches such as DFT+$U$ and DFT+DMFT, which introduce Hubbard like interaction terms while maintaining original symmetries, are commonly employed to describe strong-correlation-driven phenomena, including band-gap openings and quasiparticle mass renormalization \cite{Dudarev1998,Liechtenstein1995,kotliar2006}. Symmetry breaking comes into play by successfully capturing strong correlation effects, including band-gap opening in antiferromagnetic and paramagnetic transition metal oxides \cite{Zhang2020}, correlated behavior in perovskites \cite{Malyi2022-3}, two-dimensional materials \cite{Zunger2025}, complex phase transitions in correlated materials \cite{Joshi2024,ZiKuiLiu2024}, and strongly correlated limits in molecular systems \cite{perdew2024,Perdew2021,Perdew2022,Maniar2024}, all of this without necessarily invoking corrective methods.

Similarly, traditional coupled-cluster theory is also known to fail in regimes of strong correlation in symmetry-adapted configurations, a limitation that can be traced back to the inadequacy of a single-reference determinant to qualitatively represent the many-body wave function \cite{scuseria2015}. In such situations, the electronic state is intrinsically multiconfigurational, and improving the description would require either multireference formulations -- highly non-trivial -- or a systematic extension of the correlator toward full configuration interaction -- computationally expensive \cite{scuseria2015}. In practice, a widely adopted and effective strategy is therefore to allow for symmetry breaking at the mean-field level, which provides a qualitatively improved reference and enables a more accurate description of correlated states, by using coupled cluster theory, within feasible computational cost \cite{scuseria2015}. Nevertheless, symmetry breaking does not guarantee success in every system for coupled cluster theory \cite{LeBlanc2015}.

Symmetry breaking, however, isn't just a theoretical artifact or only just an easier path to describe strongly correlated phenomena. It is, indeed, a physical fact with experimental evidence \cite{Anderson1972,Rivera2025,Ramirez1997,sangiorgio2018}. In real systems, symmetry breaking effectively emerges due to the finite time scales over which experiments probe physical states \cite{perdew-zunger-2025,Perdew2021,Perdew2025,Anderson1972}. Even when the Hamiltonian is fully symmetric, a system may appear in a symmetry-broken configuration because of fluctuations in the expectation values of operators that do not commute with the Hamiltonian \cite{perdew-zunger-2025}. The characteristic period of these fluctuations increases with system size and, in the thermodynamic limit, can become effectively infinite if compared with any realistic experimental observation time \cite{Anderson1972,perdew-zunger-2025,Perdew2021,Perdew2025}. Magnetic states, for instance, can be understood as long-lived fluctuations in the spin density: although the exact ground state may be symmetric, the system remains trapped in one of the ordered configurations on experimentally relevant time scales \cite{perdew-zunger-2025,Perdew2021}. A fully symmetric solution therefore corresponds to an all-time average over all such configurations. However, this symmetric description may involve near-degeneracies and strong correlations, which are not optimally captured by standard approaches such as approximate density functional theory.

Based on this, the hypothesis explored in this work is that symmetry breaking in the Kohn–Sham non-interacting system, by lifting near-degeneracies around the Fermi level, effectively reduces the correlation demands of the corresponding interacting system \cite{perdew-zunger-2025}. Since near-degeneracies enhance the need for strong correlation effects, their removal through symmetry breaking transforms an otherwise strongly correlated symmetric configuration into a normally correlated one, for which density functional theory is more reliable. As a result, the symmetry-broken solution can recover a significant portion of the missing stabilization energy, leading to a substantial lowering of the total energy. Therefore, the density-functional approximated energy difference between the symmetric and symmetry-broken configurations can be interpreted as an approximation to the correlation energy associated with the strongly correlated character of the symmetric phase \cite{perdew-zunger-2025}.

Here, we investigate the main hypothesis using density functional theory \cite{kohn1964,Kohn1965} by analyzing how the number of states near the Fermi level changes upon spin symmetry breaking. Figure \ref{fig:figure0} illustrates the expected behavior. We define a correlation parameter ($\Gamma$) by comparing the density of states (DOS) of a target system at the Fermi level with that of the uniform electron gas, our normally correlated prototypical system. We relate it to $\Delta E$, the density-functional approximated total energy difference between symmetric and symmetry-broken configurations ($\Delta E = E_{symm.} - E_{symm.brok}$). Our test set includes well-known strongly correlated materials—Ni, Fe, NiO, VO$_2$ -rutile (r) and monoclinic (m) -, Co, Pd, EuB$_6$, SmB$_6$, and Gd—as well as normally correlated examples—Cu, Ag, Zn and Cd. Within this framework, $\Gamma$ serves as an indicator of DFT reliability within strongly correlated physics: $\Gamma \leq 1$ corresponds to a normally correlated regime where DFT performs well, while $\Gamma \gg 1$ signals strong correlation, where DFT becomes less reliable. In such cases, $\Gamma$ can guide whether to allow symmetry breaking within DFT or, if symmetry is to be preserved, to adopt methods that explicitly treat strong correlation. A comparisson of our results with another well established density of states based theory - the Stoner framework - is also discussed in section \ref{sec_sup:stoner} of the supplemental information.


Our test set is summarized in table \ref{tab:symmetric_SB_configurations}, together with its symmetric and symmetry broken configurations, and the density-functional approximated energy difference between them. All energy differences are normalized by the number of atoms in the respective cell. As we focus on spin symmetry, the symmetric configuration is always the nonmagnetic, while the symmetry broken state is a magnetic configuration. Symmetry broken states are chosen to be the configuration with the lowest energy for each material within the magnetic configurations landscape, determined by $r^2\mathrm{SCAN}$ as the standard density functional approximation used to model exchange and correlation effects in this work \cite{Furness2020}. To perform density functional theory calculations we used VASP \cite{vasp1,vasp2} and the postprocessing was done by using VASPKIT \cite{vaspkit}. All calculations were performed within the Projector-Augmented Wave (PAW) formalism \cite{paw}. Computational parameters are explicitly discussed in section \ref{sec_sup:numericals} of the supplemental information.

\begin{table}[h]
\centering
\caption{Symmetric (NM) and symmetry-broken (SB) states for each material. The SB state is the magnetic ground state from $r^2\mathrm{SCAN}$ calculations (e.g., A-AFM (001) denotes A-type antiferromagnetism along $\hat{z}$). $\Delta E$ is the $r^2\mathrm{SCAN}$ total energy difference between symmetric and symmetry broken states, normalized per atom. Materials with nonzero $\Delta E$ (strongly correlated) are separated from normally correlated ones ($\Delta E = 0$) by a solid line. Pd is non-magnetic in reality, but slightly magnetic in $r^2\mathrm{SCAN}$.}
\label{tab:symmetric_SB_configurations}
\begin{tabular}{lccc}
\hline
\hline
Material & Symmetric state & SB state & $\Delta E$ (eV/atom) \\
\hline
Ni              & NM & FM            & 0.116 \\
Fe              & NM & FM            & 0.909 \\
NiO             & NM & A-AFM (111)    & 0.493 \\
VO$_{2r}$ & NM & FM            & 0.071 \\
VO$_{2m}$ & NM & FM        & 0.030 \\
Co              & NM & FM            & 0.436 \\
Pd              & NM & FM            & 0.014 \\
EuB$_6$         & NM & FM            & 1.095 \\
SmB$_6$         & NM & A-AFM (001)    & 0.750 \\
Gd              & NM & FM            & 8.985 \\
\hline
Cu              & NM & NM            & 0.000 \\
Ag              & NM & NM            & 0.000 \\
Zn              & NM & NM            & 0.000 \\
Cd              & NM & NM            & 0.000 \\
\hline
\hline
\end{tabular}
\end{table}

To quantify the amount of correlation from the density of states, we define the $\Gamma$ parameter:
\begin{equation}
    \label{eq:alpha_fermi_level}
    \Gamma = \frac{\mathrm{D}(\epsilon_F)}{\mathrm{D}_{unif}(\epsilon_F)},
\end{equation}

\noindent where $\mathrm{D}(\epsilon_F)$ is the density of states at the Fermi level obtained from an $r^2\mathrm{SCAN}$ calculation, and $\mathrm{D}_{unif}(\epsilon_F)$ is the density of states at the Fermi level \cite{kittel2005introduction,ashcroft2011solid} of the normally-correlated uniform electron gas
\begin{equation}
    \label{eq:uniform_gas_DOS}
    \mathrm{D}_{unif}(\epsilon) = \frac{V_{cell}}{2\pi^2} \left( \frac{2m_e}{\hbar^2} \right)^{3/2} \sqrt{\epsilon}.
\end{equation}
Here, the Fermi level of the uniform electron gas is defined as
\begin{equation}
    \label{eq:uniform_gas_fermi_level}
    \epsilon_F = \frac{\hbar^2}{2m_e}(3 \pi^2 n)^{2/3}
\end{equation}
and set to zero from now on, so $\Gamma$ is defined as $\Gamma = \frac{\mathrm{D}(0)}{\mathrm{D}_{unif}(0)}$. In Eq. \ref{eq:uniform_gas_fermi_level} $n=N_e/V_{cell}$, with $N_e$ the number of valence correlated electrons and $V_{cell}$ is the cell volume, discussed further later. $m_e = 9.109 \times 10^{-31}\:\mathrm{kg}$ and $\hbar = 1.054 \times 10^{-34}\:\mathrm{J \cdot s}$ denote the electron mass and the reduced Planck constant, respectively, both expressed in SI units. Then,  $\mathrm{D}_{unif}(\epsilon)$ units are converted into $\frac{1}{eV}$, the same unit used to calculate $\mathrm{D}(\epsilon)$ from first principles. Within the main hypothesis, a system is considered strongly correlated when $\Gamma \gg 1$, whereas for $\Gamma \leq 1$ the material is classified as normally correlated. The current definition of $\Gamma$ is chosen to be the simplest one -- the rough density of states at the Fermi level. It is worth remembering that strong correlation requires nearly degenerate \textit{occupied and unoccupied states}. In section \ref{sec_sup:alphas}, we present a generalization of Eq. \ref{eq:alpha_fermi_level}, the Gaussian-smeared $\Gamma_g$, which requires an energy width $\delta$ around the Fermi level and reduces to Eq. \ref{eq:alpha_fermi_level} as $\delta \rightarrow 0$. Although the value of $\delta$ for a given class of materials (e.g., the 3d transition elements and their compounds) is fitted, the resulting $\Gamma_g$ seems to be much more robustly predictive of strong correlation than $\Gamma$ of Eq. \ref{eq:alpha_fermi_level} is. We discuss this in detail on section \ref{sec_sup:alphas} of the supplemental information, also taking care of asymmetries between the number of unoccupied and occupied states.

In the uniform electron gas model, there are only two parameters: the cell volume $V_{cell}$ and the number of correlated electrons $N_e$. To ensure a consistent comparison with \textit{ab initio} calculations, we always use the same cell volume employed in the first-principles computations, along with the number of correlated valence electrons relevant to each material. The cell volume $V_{\text{cell}}$ is obtained by fully relaxing the structure in its symmetric configuration, ensuring that only spin-symmetry breaking is considered, without allowing any symmetry breaking in the atomic positions. For $N_e$, we always use the total number of correlated valence electrons within the cell of volume $V_{cell}$. For instance, there are 2 Fe atoms in the unit cell and 7 valence \textit{d} electrons for each Fe atom, totalizing 14 \textit{d} correlated valence electrons in the system ($N_e = 14$). Details on the cell volumes, symmetries and number of valence electrons are given in section \ref{sec_sup:numericals} of the supplemental information.

\begin{figure}[h!]
    \centering
    \includegraphics[width=0.85\linewidth]{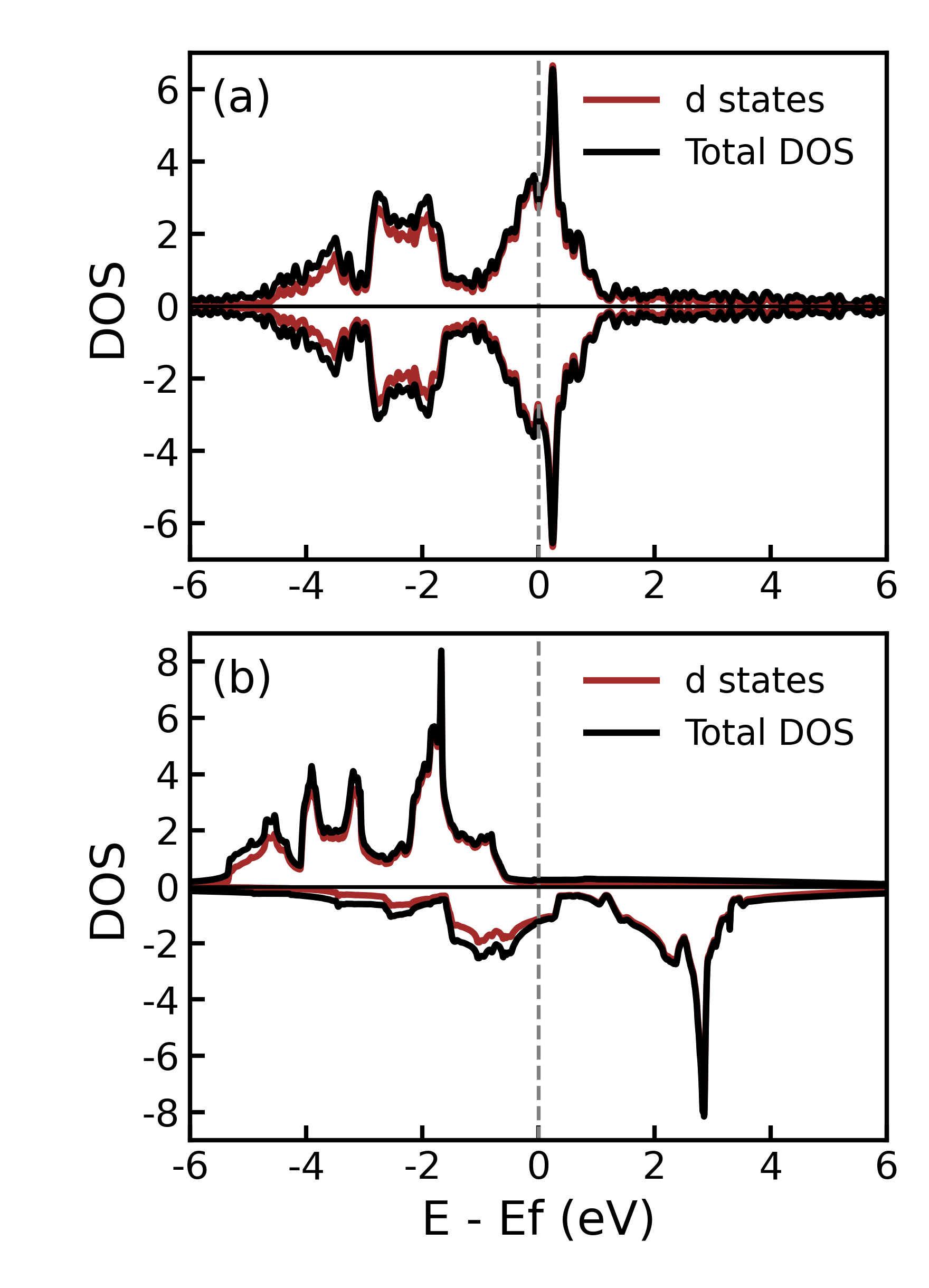}
    \caption{Densities of states for (a) symmetric (nonmagnetic) and (b) symmetry-broken (ferromagnetic) Fe. The density of states is plotted upward for spin $\uparrow$ and downward for spin $\downarrow$.}
    \label{fig:Fe_dos}
\end{figure}
\begin{figure}[h!]
    \centering
    \includegraphics[width=0.85\linewidth]{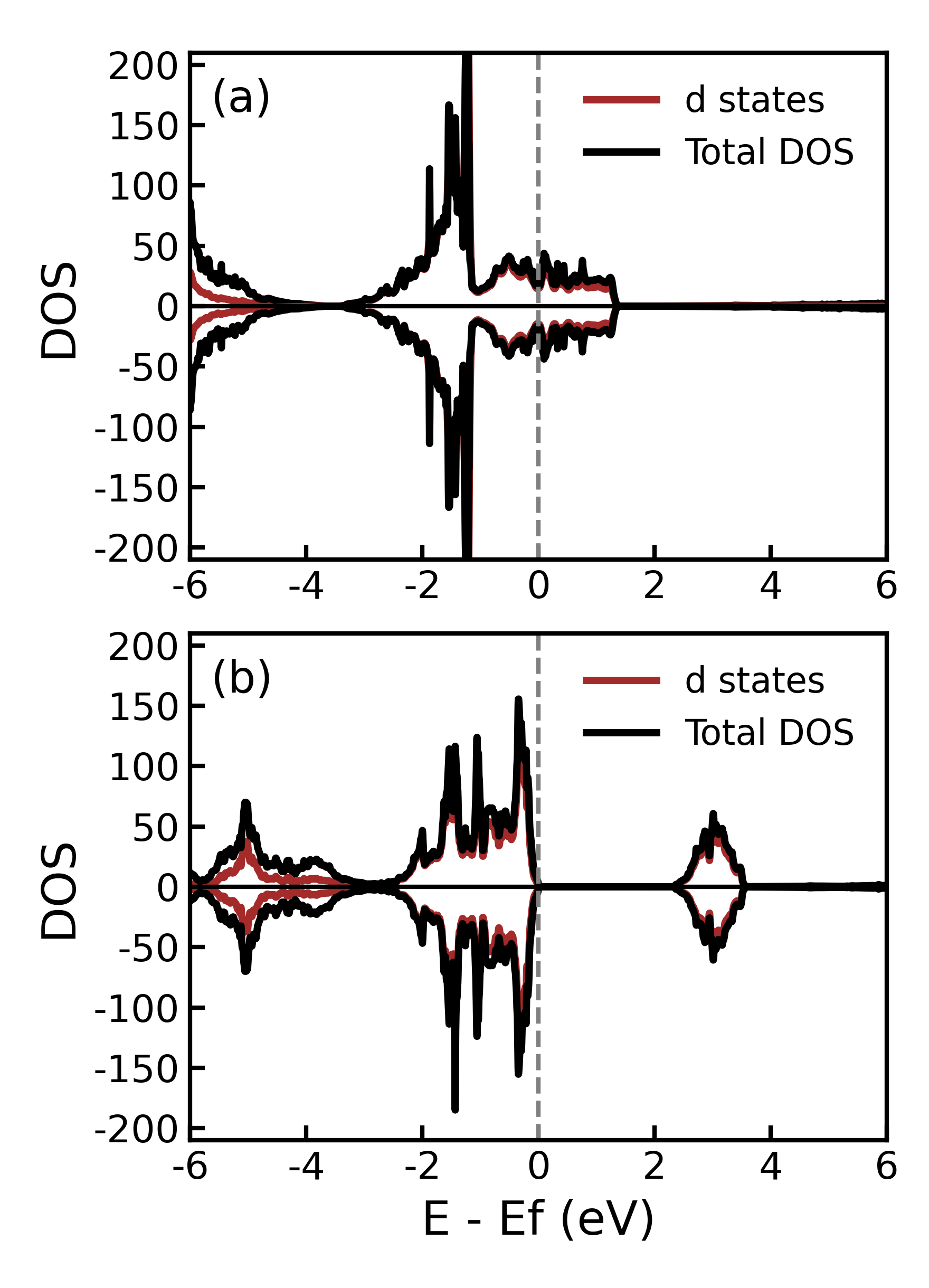}
    \caption{Densities of states for (a) symmetric (nonmagnetic) and (b) symmetry-broken (antiferromagnetic) NiO. The density of states is plotted upward for spin $\uparrow$ and downward for spin $\downarrow$.}
    \label{fig:NiO_dos}
\end{figure}

We plot the density of states for Fe and NiO, figures \ref{fig:Fe_dos} and \ref{fig:NiO_dos} respectively, for both symmetric and symmetry broken configurations, which agrees with previous calculations \cite{mohn2002magnetism,Zhang2020,DelloStritto2023}. The density of states for each material within our test set is shown in section \ref{sec_sup:dos} of the supplemental material. Clearly we can see that, after symmetry breaking, the number of states near the Fermi level decreases considerably for both systems. There is actually a band gap opening in NiO consistent with experimental and other theoretical results \cite{Sawatzky1984,Zhang2020,DelloStritto2023}.

Under the assumption that symmetry breaking can transform a symmetric strongly correlated configuration into a normally correlated one, we propose the typical—though not necessarily universal—behavior. In a symmetric many-electron state, the density of one-electron states associated with localized $d$ or $f$ electrons at the Fermi level is substantially higher than in the broken-symmetry configurations. As a consequence, the symmetric state is strongly correlated and, within density functional theory calculations, tends to be unstable with respect to the formation of a symmetry-broken many-electron state, which is characterized by a reduced density of states at the Fermi level, more negative total energy and a more weakly correlated electronic structure.

This symmetry breaking may, in particular for Mott materials such as NiO, lead to an insulating state with a finite fundamental gap. Moreover, a given symmetry-broken configuration may itself be only marginally stable and may further evolve into another symmetry-broken state with a lower density of states at the Fermi level. When the density of states at the Fermi level becomes sufficiently small, or when the band gap becomes sufficiently large, strong correlation effects and the driving force for symmetry breaking may be effectively suppressed, indicating that the system has reached a regime that can be described as normally correlated.

\begin{figure}[h!]
    \centering
    \includegraphics[width=0.85\linewidth]{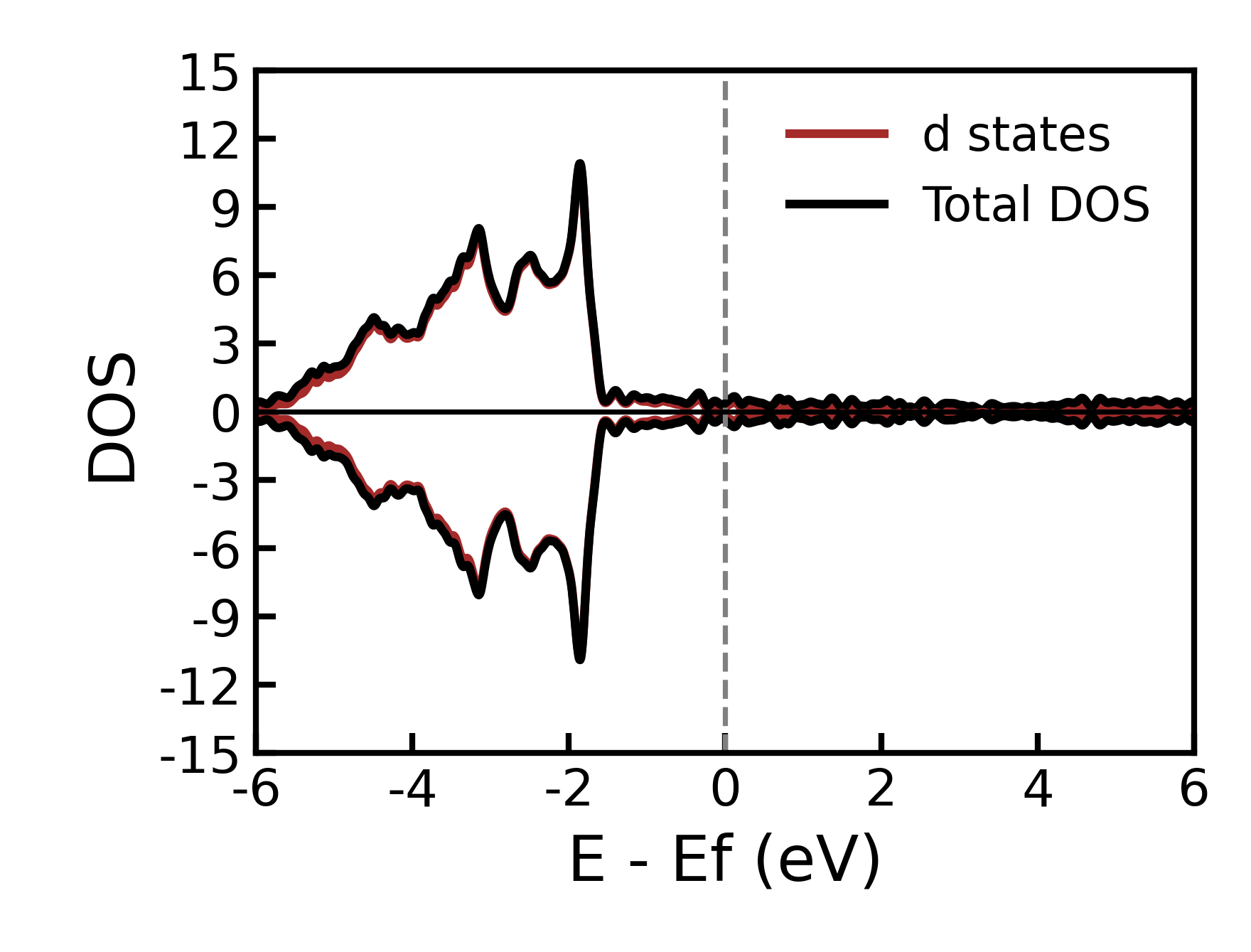}
    \caption{Density of states for symmetric (nonmagnetic) Cu.}
    \label{fig:Cu_dos}
\end{figure}

As we take a look at table \ref{tab:DOS_ratios}, symmetry breaking did not affect any of the normally correlated examples. In other words, when breaking spin symmetry, all normally correlated materials converged to a nonmagnetic configuration, or the symmetric state, resulting in $\Delta E = 0$. Furthermore, these systems show a considerably low density of states in the vicinity of the Fermi level in the symmetric phase - which results in low $\Gamma$ values -, indicating normal correlation within the main hypothesis. This fact is exemplified by Cu in figure \ref{fig:Cu_dos}, and can be seen for the other normally correlated examples in section \ref{sec_sup:dos} of the supplemental information.

By further analyzing the quantitative values of $\Gamma$ in Table \ref{tab:DOS_ratios}, we observe that $\Gamma$ clearly differentiates between strongly and normally correlated materials within our dataset: when $\Delta E = 0$, \textit{i.e.}, when symmetry is not broken, $\Gamma \le 1$, and when $\Delta E \neq 0$ and symmetry breaks, $\Gamma > 1$. 
This indicates that both descriptors qualitatively agree on the classification of correlated behavior. However, as already discussed, there are intrinsic limitation to this approach: we rely on the Kohn–Sham non-interacting system, described by an approximate exchange–correlation functional tailored to normally correlated systems and with self-interaction pathologies \cite{SIC_perdew_zunger,Perdew2025}, to represent a symmetric state that may become strongly correlated in the interacting picture. As a result, the density of states near the Fermi level may not be quantitatively accurate, which likely explains why $\Gamma$ provides a qualitative, rather than fully quantitative, indicator of correlation.
\begin{table}[h!]
\centering
\caption{Fermi-level density of states ratios, and the corresponding $r^2\mathrm{SCAN}$ energy difference $\Delta E$ between SB and NM states.}
\label{tab:DOS_ratios}
\begin{tabular}{lccc}
\hline
\hline
Material &
$\Gamma$ &
$\mathrm{D}_{\mathrm{NM}}(0)/\mathrm{D}_{\mathrm{SB}}(0)$ &
$\Delta E$ (eV/atom) \\
\hline
Ni   & 9.140 & 2.838 & 0.116 \\
Fe   & 7.222  & 3.090 & 0.909 \\
NiO  & 1.903  & $\infty$ & 0.493 \\
VO$_2$ (r)     & 4.640 & 2.255 & 0.071 \\
VO$_2$ (m) & 1.881 & 2.656 & 0.030 \\
Co   & 4.949 & 2.763 & 0.436 \\
Pd   & 4.407 & 1.631 & 0.014 \\
EuB$_6$ & 6.681 & 4.145 & 1.095 \\
SmB$_6$ & 3.879 & 2.923 & 0.750 \\
Gd   & 23.074 & 33.667 & 8.985 \\
\hline
Cu   & 0.667 & 1.000 & 0.000 \\
Ag   & 0.526 & 1.000 & 0.000 \\
Zn   & 0.702 & 1.000 & 0.000 \\
Cd   & 0.391 & 1.000 & 0.000 \\
\hline
\hline
\end{tabular}
\end{table}

\begin{figure}[h!]
    \centering
    \includegraphics[width=1\linewidth]{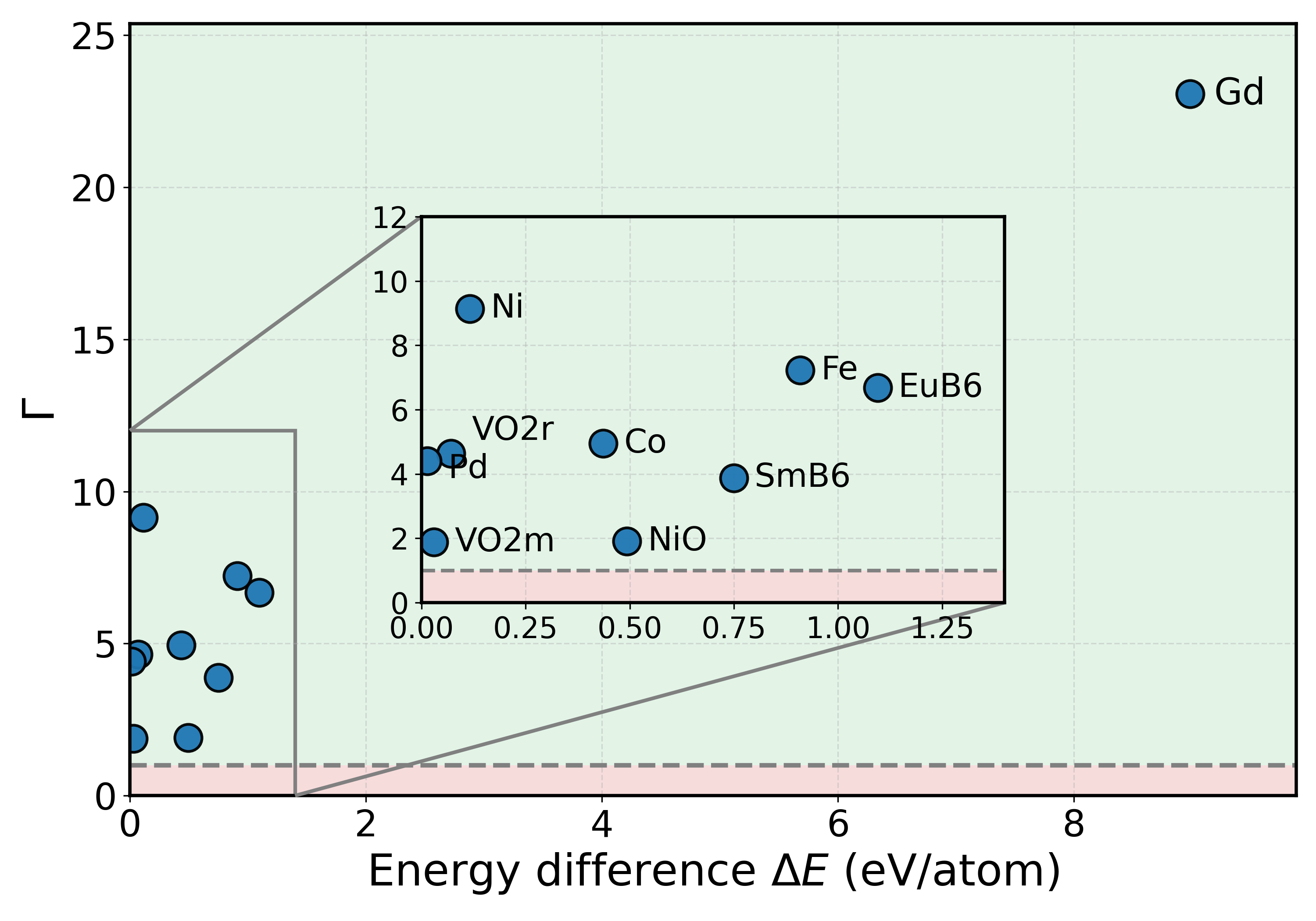}
    \caption{$\Delta E$ \textit{vs} $\Gamma$ plot. Green (red) region is for strongly (normally) correlated system. 
    }
    \label{fig:alpha_vs_deltaE}
\end{figure}

Focusing on the density of states ratio $\mathrm{D}{NM}(0)/\mathrm{D}{SB}(0)$, when $\mathrm{D}{NM}(0)/\mathrm{D}{SB}(0) = 1$, it means the material converged to a symmetric state even when nudging symmetry breaking, i.e., initializing the calculation from a state with explicitly broken spin symmetry, with nonzero local magnetic moments assigned to each site. The ratio tends to infinity for NiO since symmetry breaking turns it into a gapped system. $\mathrm{D}_{NM}(0)/\mathrm{D}_{SB}(0)$ also distinguishes strongly from normally correlated systems. We can see a trend between $\Gamma$ and $\mathrm{D}_{NM}(0)/\mathrm{D}_{SB}(0)$, indicating both descriptors agree qualitatively. We plot $\Delta E$ as a function of $\Gamma$ for the strongly correlated systems in figure \ref{fig:alpha_vs_deltaE}. We restrict the analysis to strongly correlated materials to compare systems governed by the same physics, as normally correlated states belong to a different regime.

By analysing Fig.~\ref{fig:alpha_vs_deltaE}, a linear dependence between $\Delta E$ and $\Gamma$ can be observed, with Ni and NiO being clear outliers. By computing the Pearson correlation coefficient, we obtain a value of 0.93; however, this result is dominated by the behavior of Gd. When Gd is excluded from the analysis, the correlation drops to 0.23, indicating a weak linear dependence between $\Delta E$ and $\Gamma$. 
However, as discussed, this behavior is strongly influenced by the number of states considered in the vicinity of $\epsilon_F$. When a larger manifold of states near $\epsilon_F$, via $\Gamma_g$ of Eq. \ref{eq:alpha_2}, is taken into account, the dependence between $\Delta E$ and $\Gamma$ becomes significantly more linear across all materials, as shown in the results on section \ref{sec_sup:alphas} of the supplemental information. For Ni, for instance, figure \ref{fig:dos_Ni} shows a big unbalanced number of occupied and unoccupied states, that when taken into account, linearized Ni behavior.

Pd constitutes an important case to the trends discussed here: although it is experimentally observed in a nonmagnetic symmetric configuration under ambient conditions, it breaks symmetry in PBE-level calculations, agreeing with our results \cite{Tran2012}. Pd is a prototypical nearly ferromagnetic metal, lying very close to satisfying the Stoner criterion \cite{mohn2002magnetism}, and it is known experimentally that the symmetric state of Pd exhibits the presence of \textit{paramagnons}, which are common signatures of nonmagnetic metals at the verge of a ferromagnetic instability \cite{doubble2010}. In parallel, recent works on VO$_2$ have shown that the dimerization present in both rutile and monoclinic phases plays an important role in attenuating magnetic symmetry breaking in these materials, thereby suppressing the tendency toward magnetic ordering \cite{Zhang2024}. Since we are not breaking positional symmetry, but we are breaking spin symmetry, this makes both VO$_2$ good cases to study spin symmetry breaking without positional symmetry breaking.



\textit{Summarizing}, symmetry breaking within standard DFT can qualitatively capture the energetic effects of strong electronic correlations by lifting near-degeneracies around the Fermi level in the Kohn–Sham non-interacting system, thereby stabilizing a configuration that corresponds to a normally correlated regime in the associated interacting system. To make this hypothesis operational, we introduced the correlation parameter $\Gamma$, defined from the density of states at the Fermi level relative to that of the uniform electron gas, which, although not containing the full physics of strong correlation, provides a practical and computable indicator within the DFT framework. In practice, $\Gamma$ can be used to assess the reliability of the approximate DFT description of the KS system: small values ($\Gamma < 1$) usually indicate that DFT is expected to perform well, while large values ($\Gamma > 1$) signal that the system may be strongly correlated, suggesting a methodological choice—either to allow symmetry breaking and rely on approximate DFT, or to maintain symmetry while explicitly treating strong correlation effects. The parameter $\Gamma$ was able to clearly distinguish strongly from normally correlated metallic systems in our dataset. Moreover, we found a trend between $\Gamma$ and the ratio $D_{\mathrm{NM}}(0)/D_{\mathrm{SB}}(0)$, indicating that systems identified as strongly correlated in their symmetric configurations are effectively transformed, through symmetry breaking in the KS system, into normally correlated ones in the corresponding interacting picture, consistent with the definition of $\Gamma$.

\textit{Aknowledgements} - The work of D.D.R. and G.M.D. was supported by agencies FAPESP (processes 2023/03493-0, 2025/08647-0 and 2023/09820-2) and CNPq. The work of J.P.P. was supported by the National Science Foundation under Grants DMR-2426275 and CHE-2533416. We thank CENAPAD-SP and LNCC (Santos Dumont Supercomputer) for computer time.



\putbib
\end{bibunit}

\begin{bibunit}

\pagebreak
\clearpage
\onecolumngrid   

\setcounter{section}{0}
\setcounter{figure}{0}
\setcounter{table}{0}
\setcounter{equation}{0}

\renewcommand{\thesection}{S\arabic{section}}
\renewcommand{\thefigure}{S\arabic{figure}}
\renewcommand{\thetable}{S\arabic{table}}
\renewcommand{\theequation}{S\arabic{equation}}

\begin{center}
{\large \bf Supplemental Information: Identifying strong correlation using only the Kohn-Sham density of one-electron states}

\vspace{0.5cm}

Daniel D. Rivera,$^{1,2}$\: Gustavo M. Dalpian,$^{2}$\: John P. Perdew$^{1}$

\vspace{0.3cm}

{\small
$^{1}$ \textit{Department of Physics and Engineering Physics, Tulane University, New Orleans, LA 70118, USA}\\
}
$^{2}$ \textit{Instituto de Física, Universidade de São Paulo, 05315-970 São Paulo, São Paulo, Brasil} 

\end{center}

\section{Densities of states}
\label{sec_sup:dos}

In this section we show every density of states not shown in the main text but calculated for the set of materials in this paper. We begin by showing the density of states for the strongly correlated examples, \textit{i.e.} the ones that actually undergo the process of symmetry breaking, followed by the normally correlated examples, or systems that do not undergo a symmetry breaking process. We can clearly see how symmetry breaking leads to a clear decrease in the number of states in the vicinity of the Fermi level for the materials in our test set.

When symmetry breaking results in a single spin channel dominating near the Fermi level, the system can be effectively interpreted as transitioning from a strongly correlated to a normally correlated regime. This is because the Pauli exclusion principle enforces spatial separation between electrons with the same spin, thereby reducing the effective Coulomb electron–electron repulsion and diminishing the need for a strong-correlation treatment.
\begin{figure}[h]
    \centering
    \includegraphics[width=0.7\linewidth]{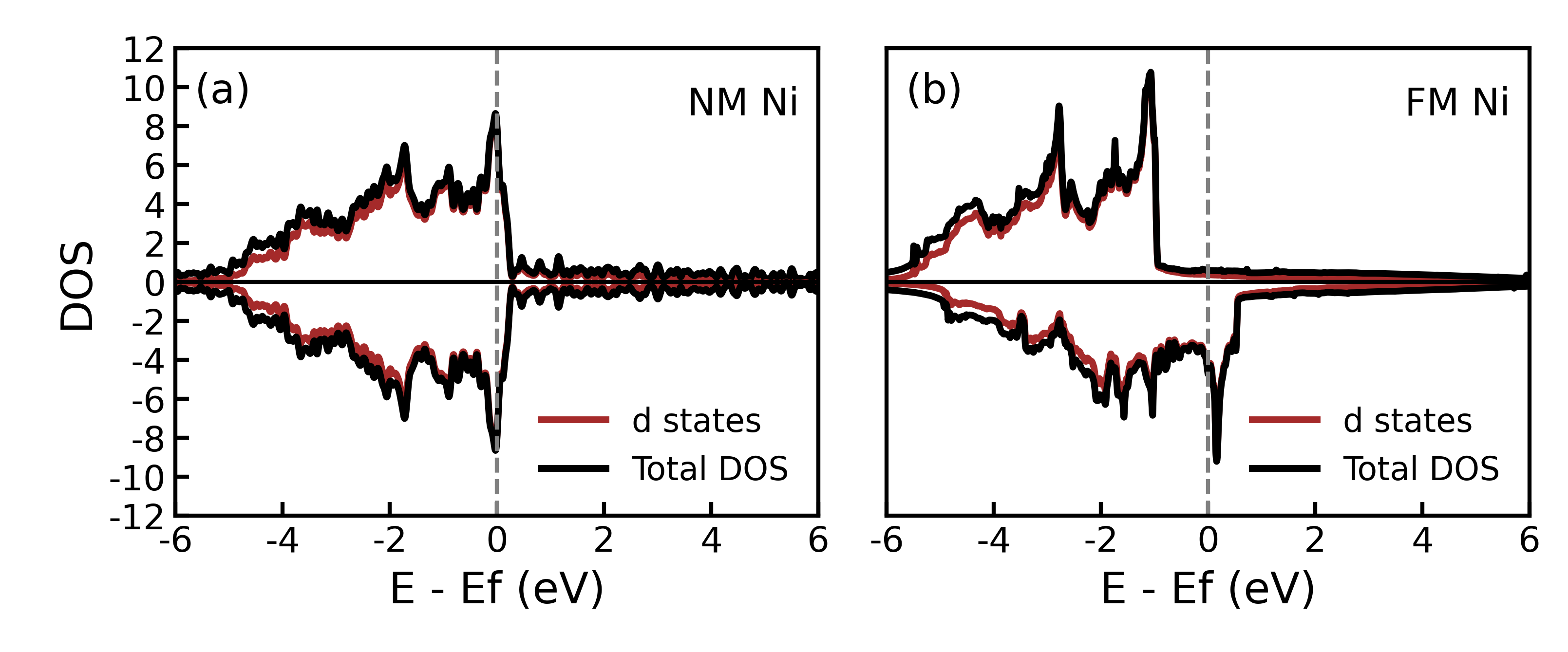}
    \caption{Ni DOS (a) before breaking symmetry, and (b) after breaking symmetry.}
    \label{fig:dos_Ni}
\end{figure}
\begin{figure}[h]
    \centering
    \includegraphics[width=0.7\linewidth]{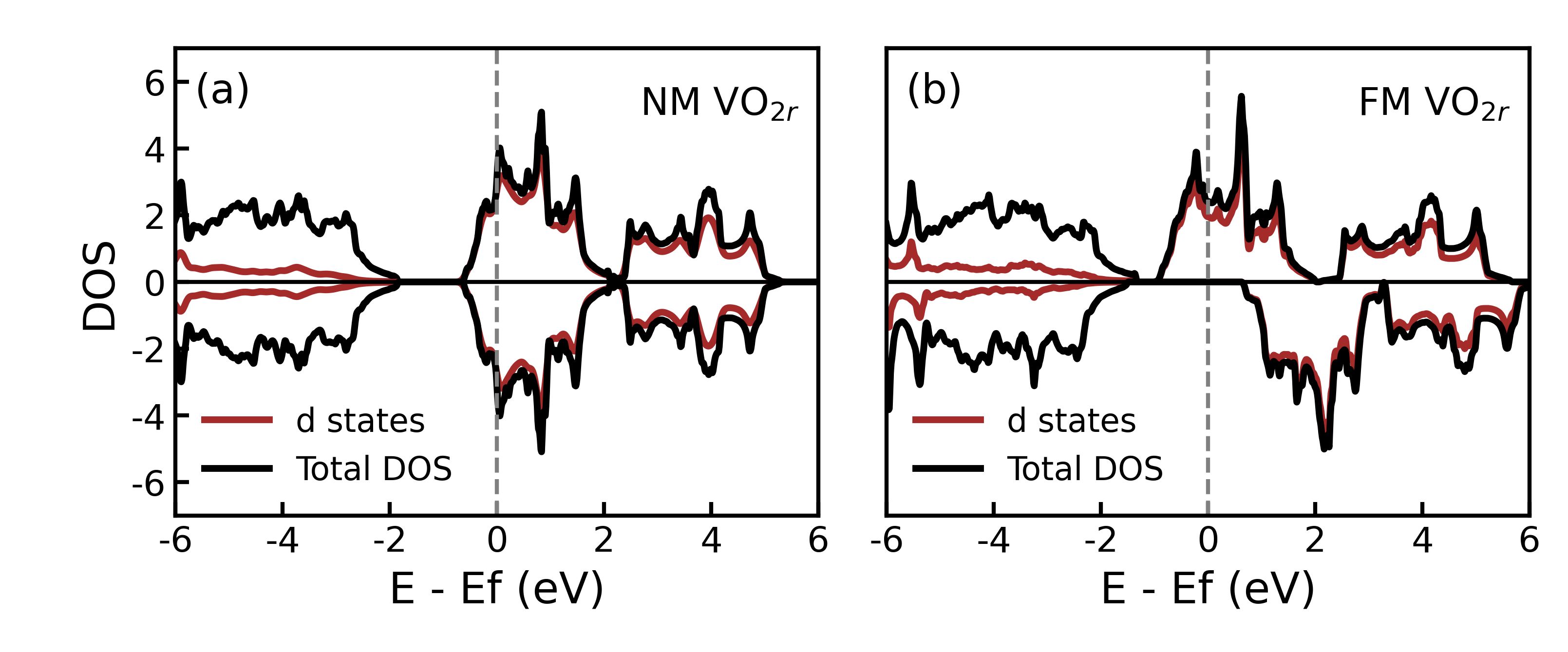}
    \caption{VO$_2$ rutile DOS (a) before breaking symmetry, and (b) after breaking symmetry.}
    \label{fig:dos_VO2r}
\end{figure}


\begin{figure}[H]
    \centering
    \includegraphics[width=0.7\linewidth]{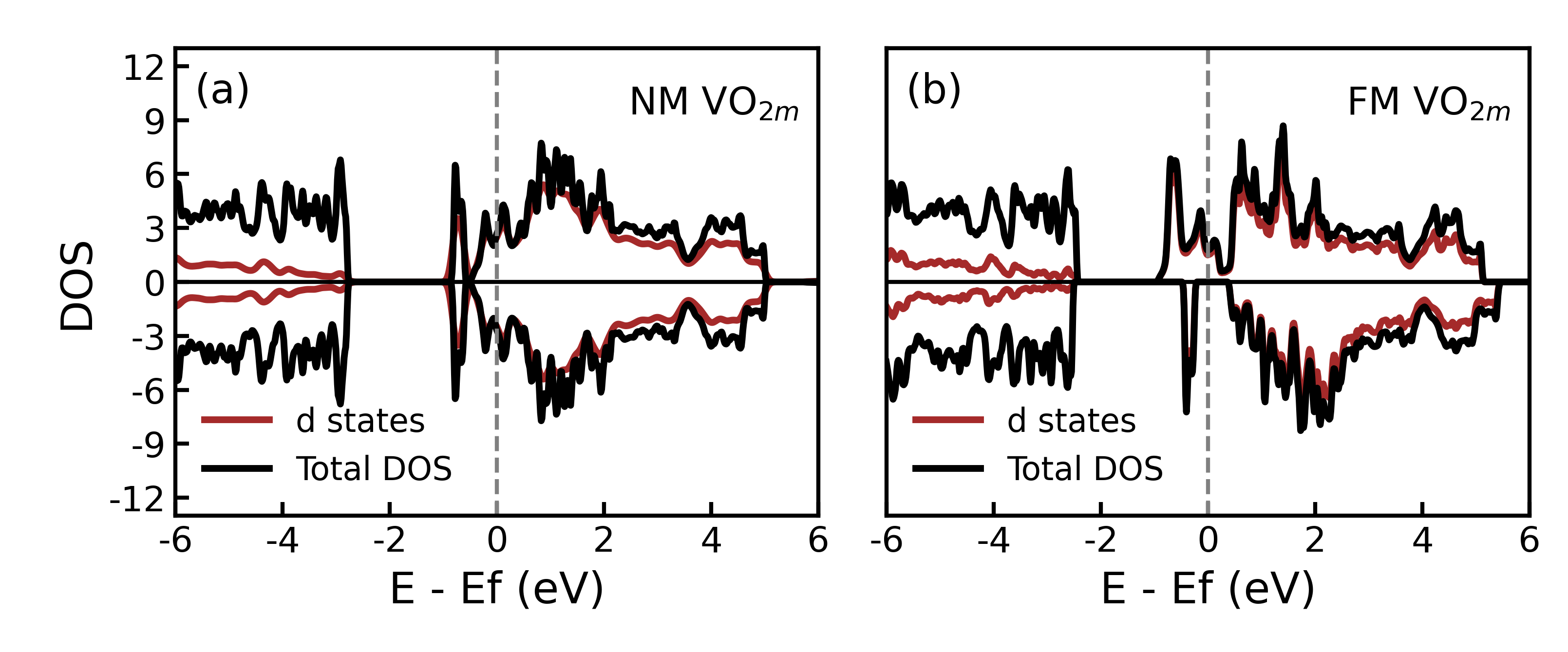}
    \caption{VO$_2$ monoclinic DOS (a) before breaking symmetry, and (b) after breaking symmetry.}
    \label{fig:dos_VO2m}
\end{figure}
\begin{figure}[H]
    \centering
    \includegraphics[width=0.7\linewidth]{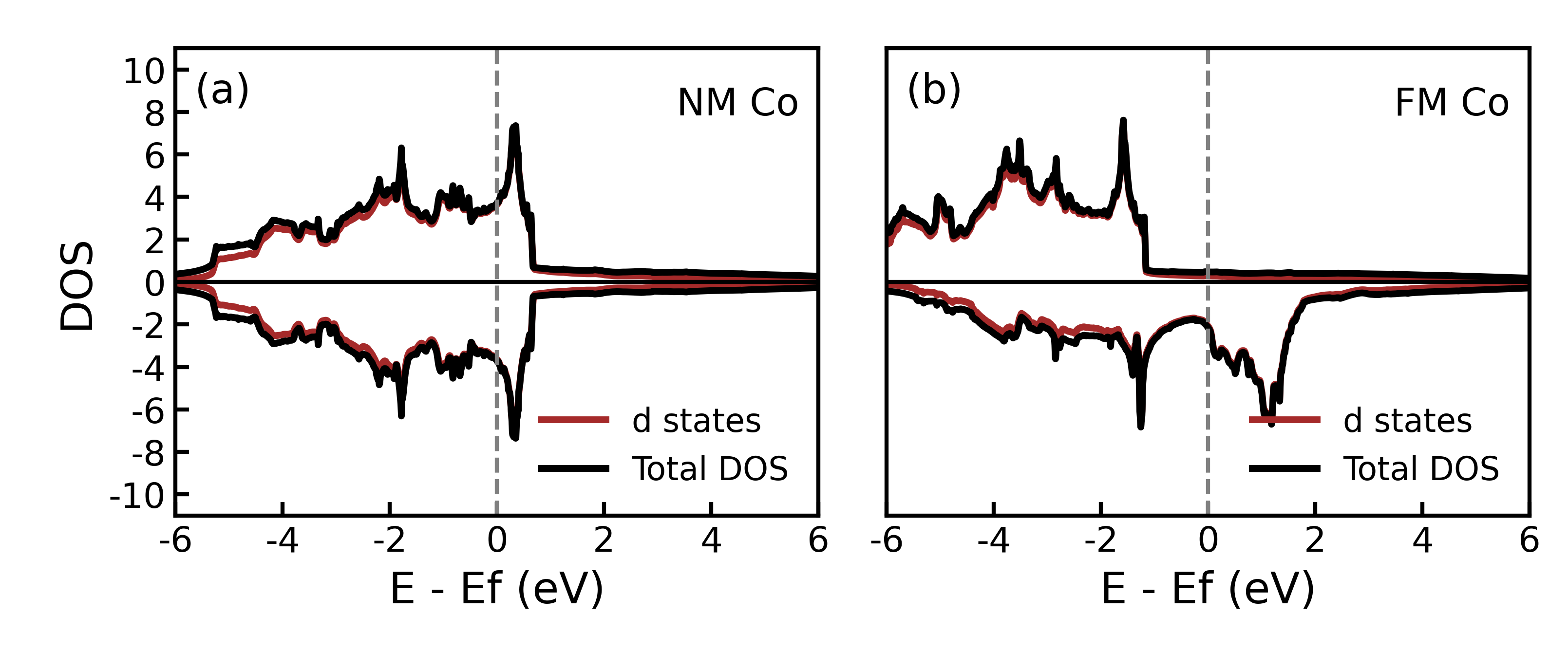}
    \caption{Co DOS (a) before breaking symmetry, and (b) after breaking symmetry.}
    \label{fig:dos_Co}
\end{figure}
\begin{figure}[H]
    \centering
    \includegraphics[width=0.7\linewidth]{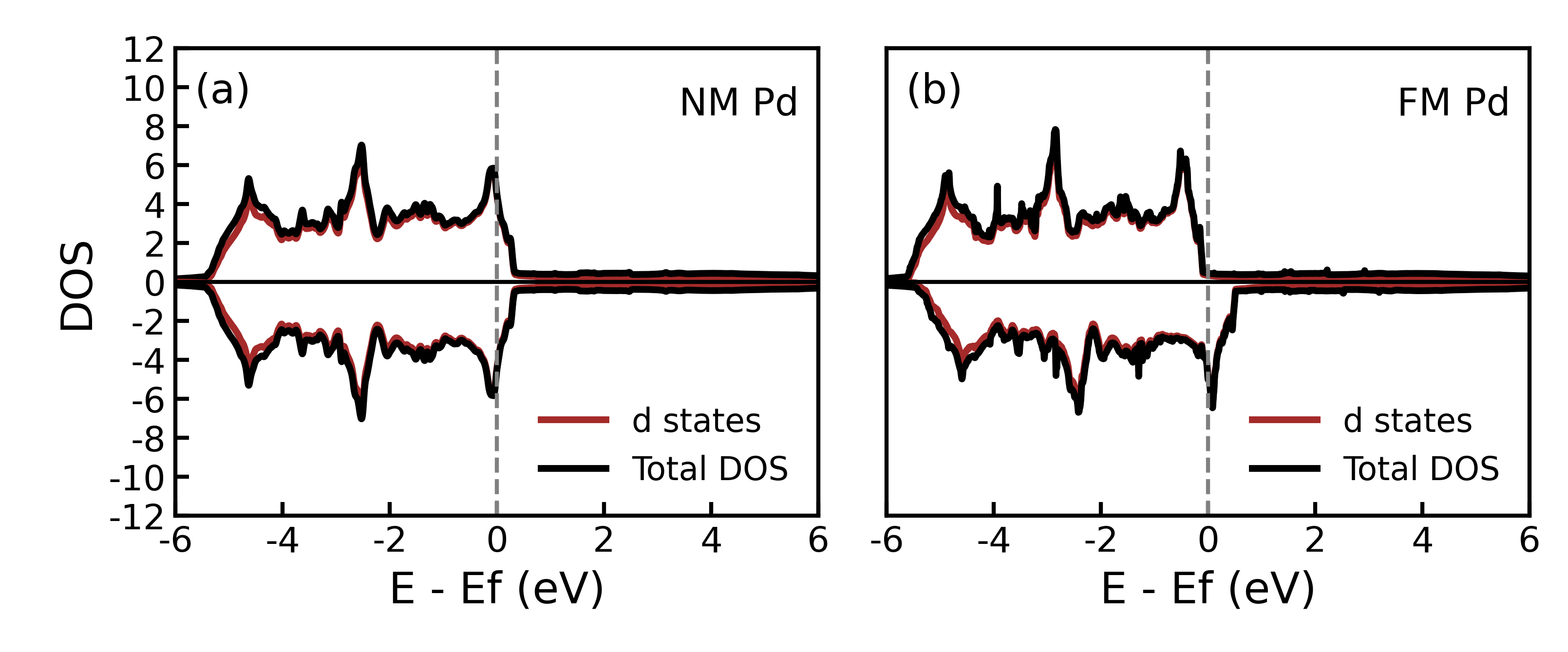}
    \caption{Pd DOS (a) before breaking symmetry, and (b) after breaking symmetry.}
    \label{fig:dos_Pd}
\end{figure}
\begin{figure}[h]
    \centering
    \includegraphics[width=0.7\linewidth]{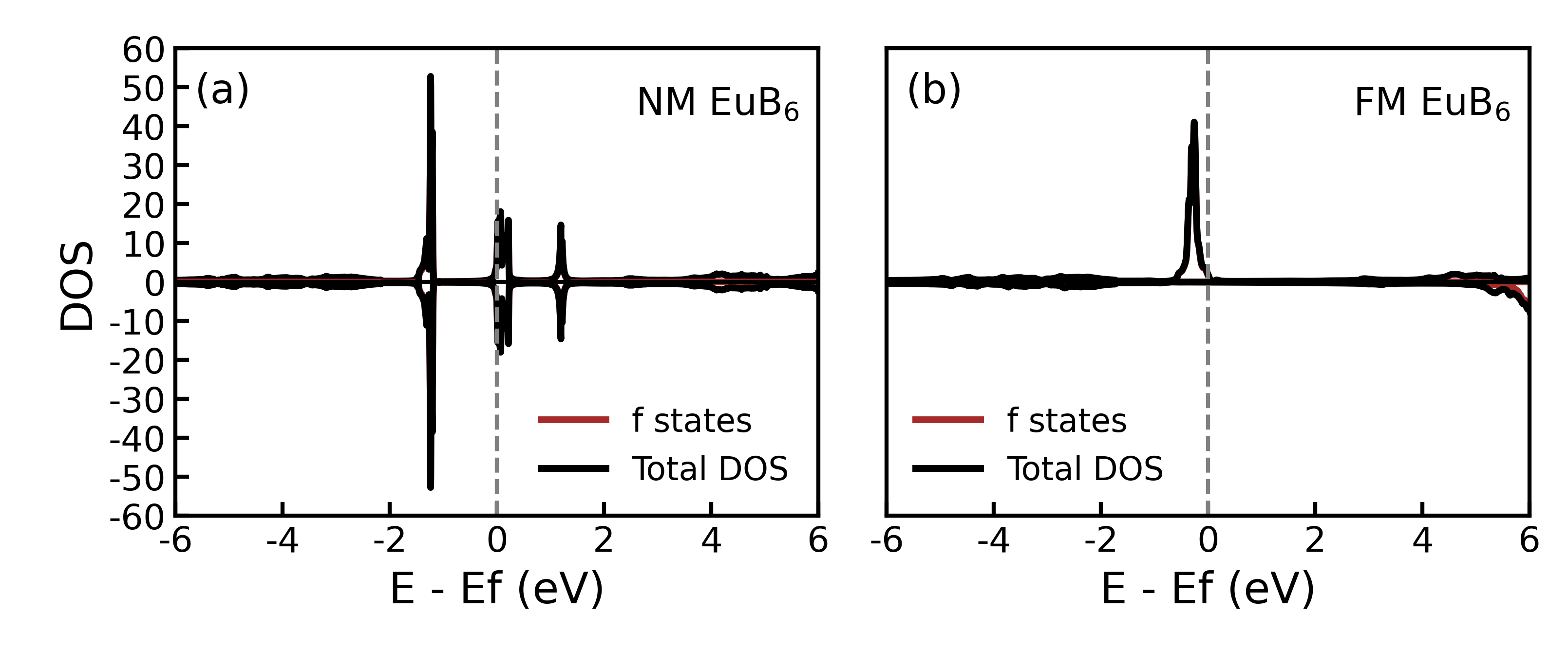}
    \caption{EuB$_6$ DOS (a) before breaking symmetry, and (b) after breaking symmetry.}
    \label{fig:dos_EuB6}
\end{figure}
\begin{figure}[h]
    \centering
    \includegraphics[width=0.7\linewidth]{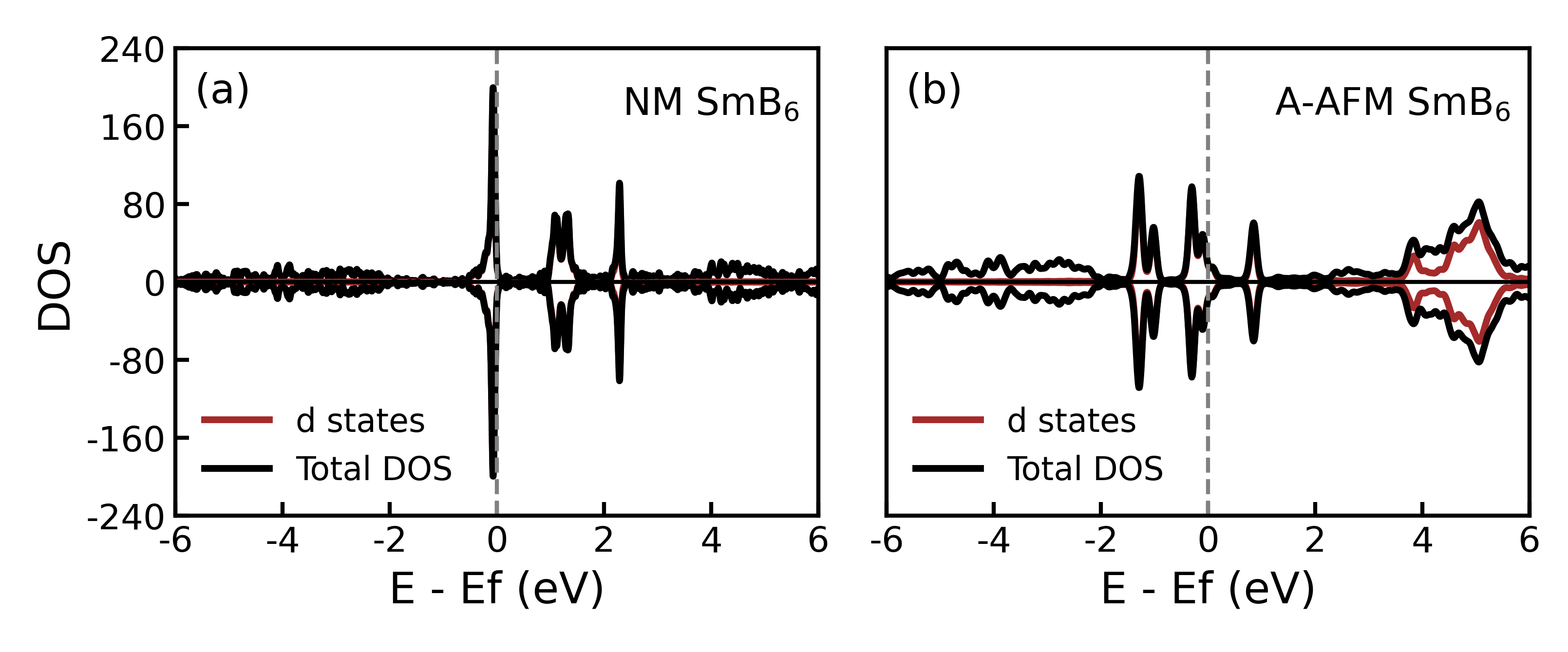}
    \caption{SmB$_6$ DOS (a) before breaking symmetry, and (b) after breaking symmetry.}
    \label{fig:dos_SmB6}
\end{figure}
\begin{figure}[h]
    \centering
    \includegraphics[width=0.7\linewidth]{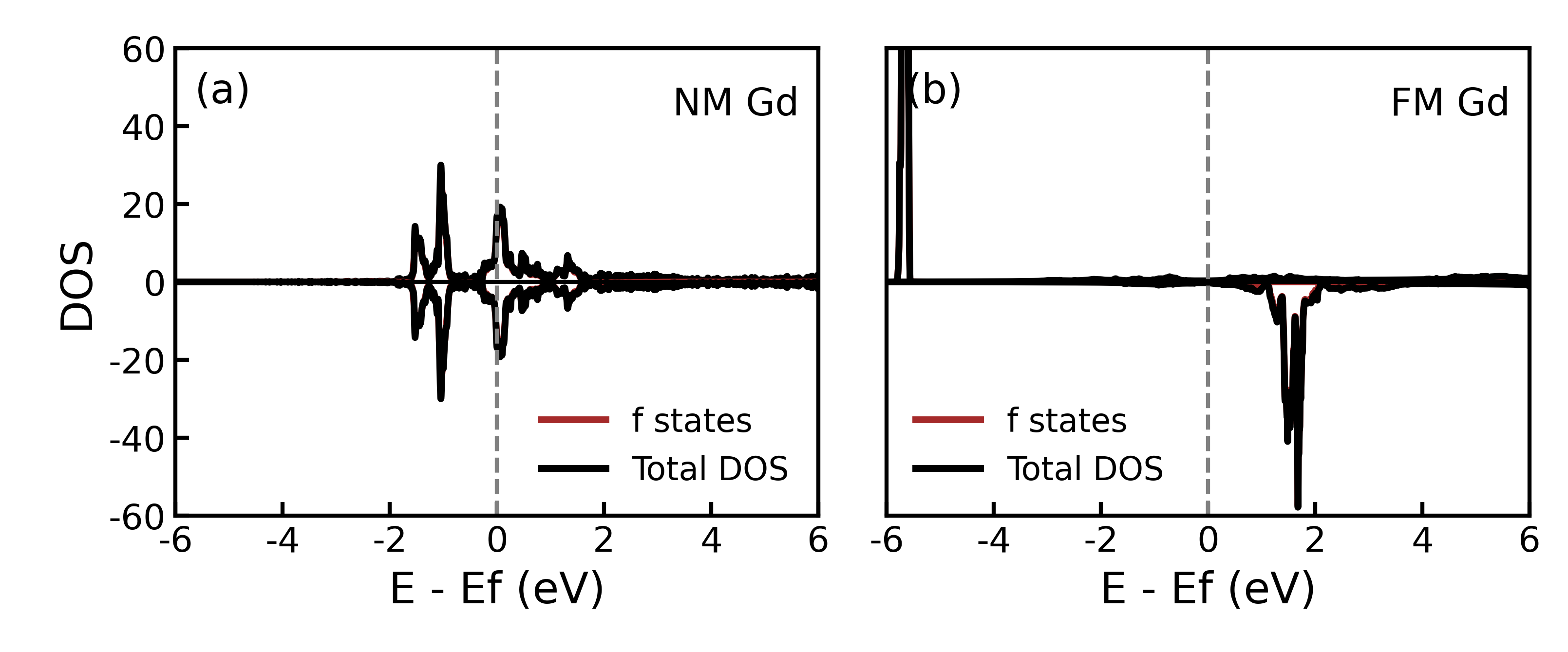}
    \caption{Gd DOS (a) before breaking symmetry, and (b) after breaking symmetry.}
    \label{fig:dos_Gd}
\end{figure}

\begin{figure}[h]
    \centering
    \includegraphics[width=0.4\linewidth]{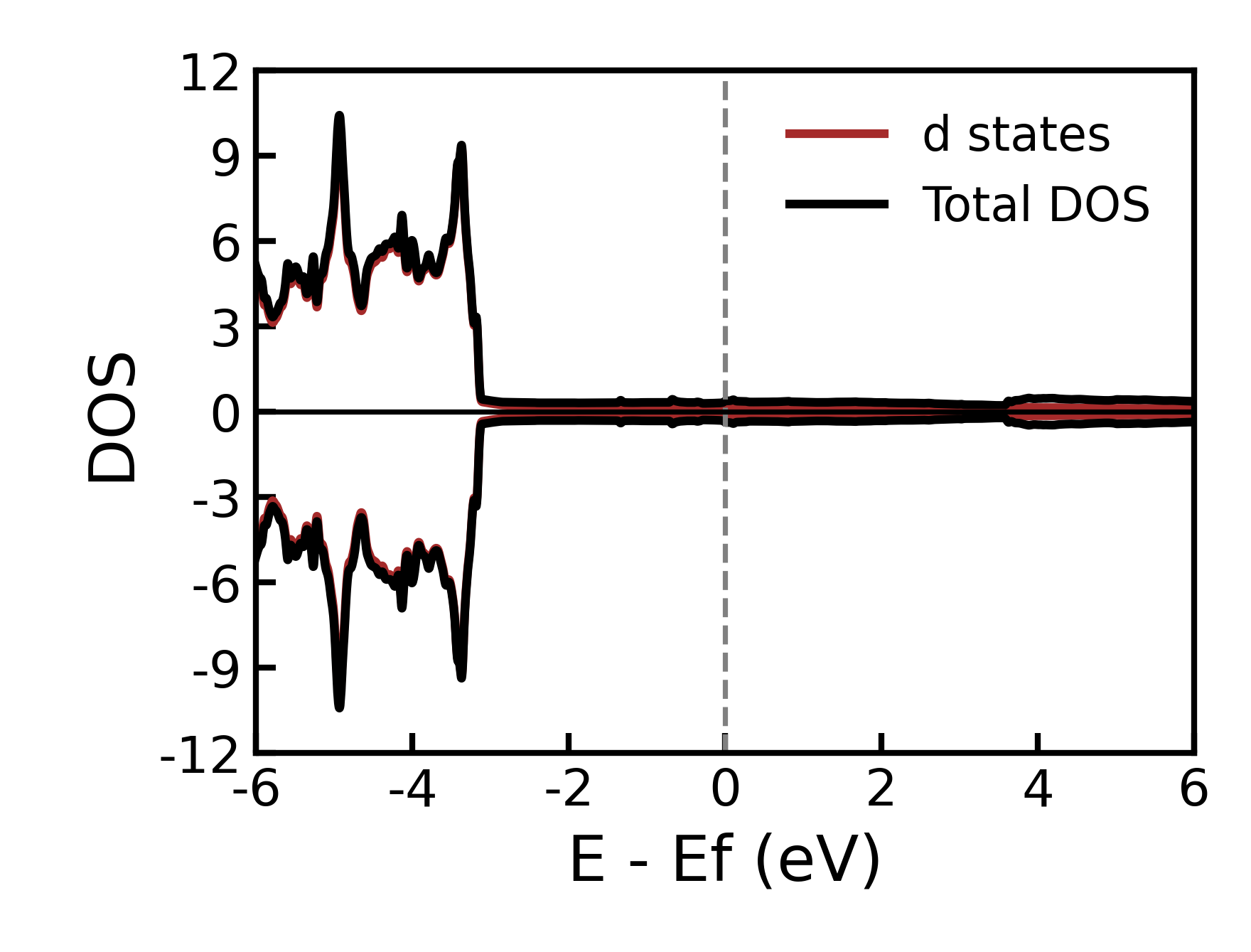}
    \caption{Ag DOS before and after nudging spin symmetry breaking.}
    \label{fig:dos_Ag}
\end{figure}
\begin{figure}[h]
    \centering
    \includegraphics[width=0.4\linewidth]{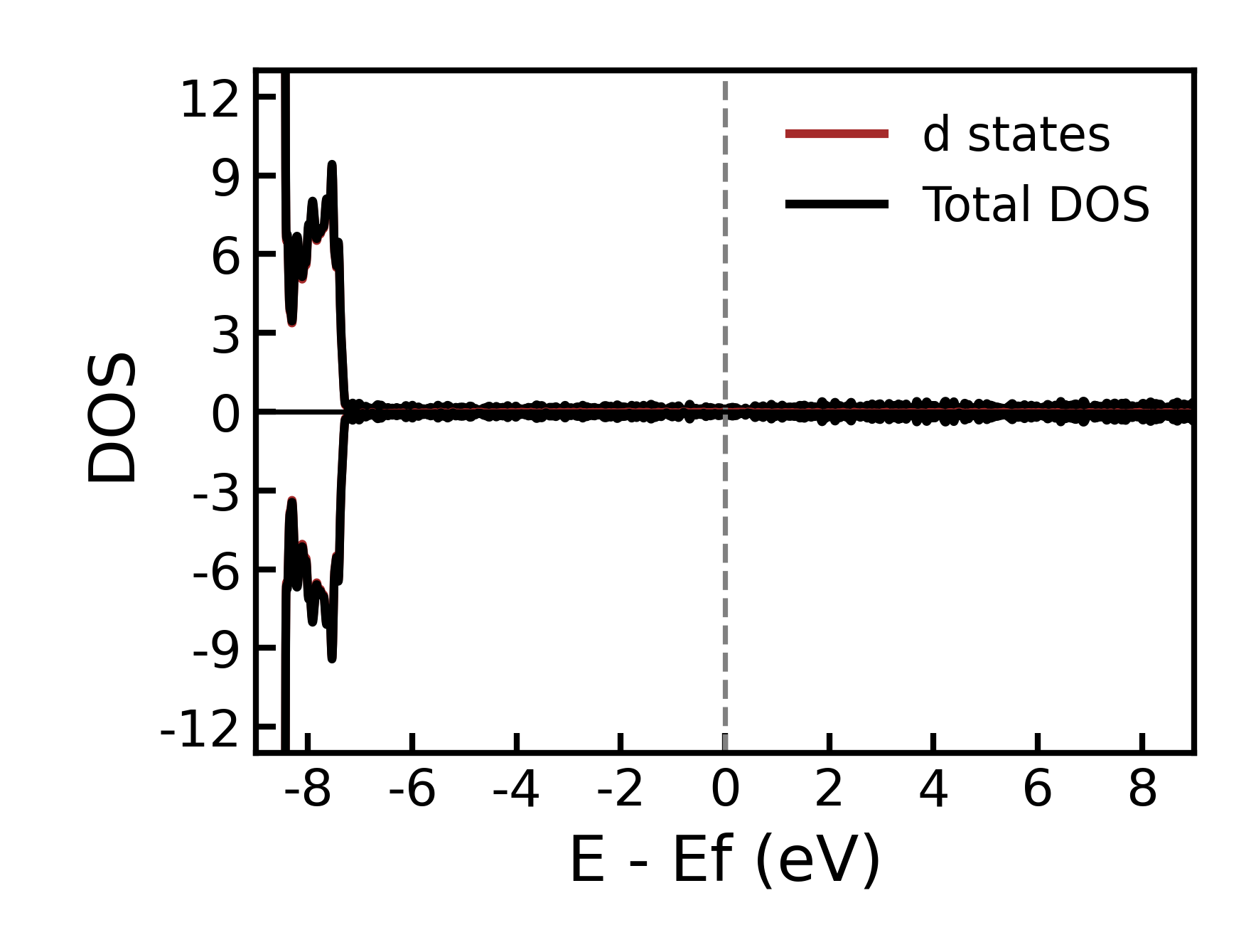}
    \caption{Zn DOS before and after nudging spin symmetry breaking.}
    \label{fig:dos_Gd}
\end{figure}
\begin{figure}[h]
    \centering
    \includegraphics[width=0.4\linewidth]{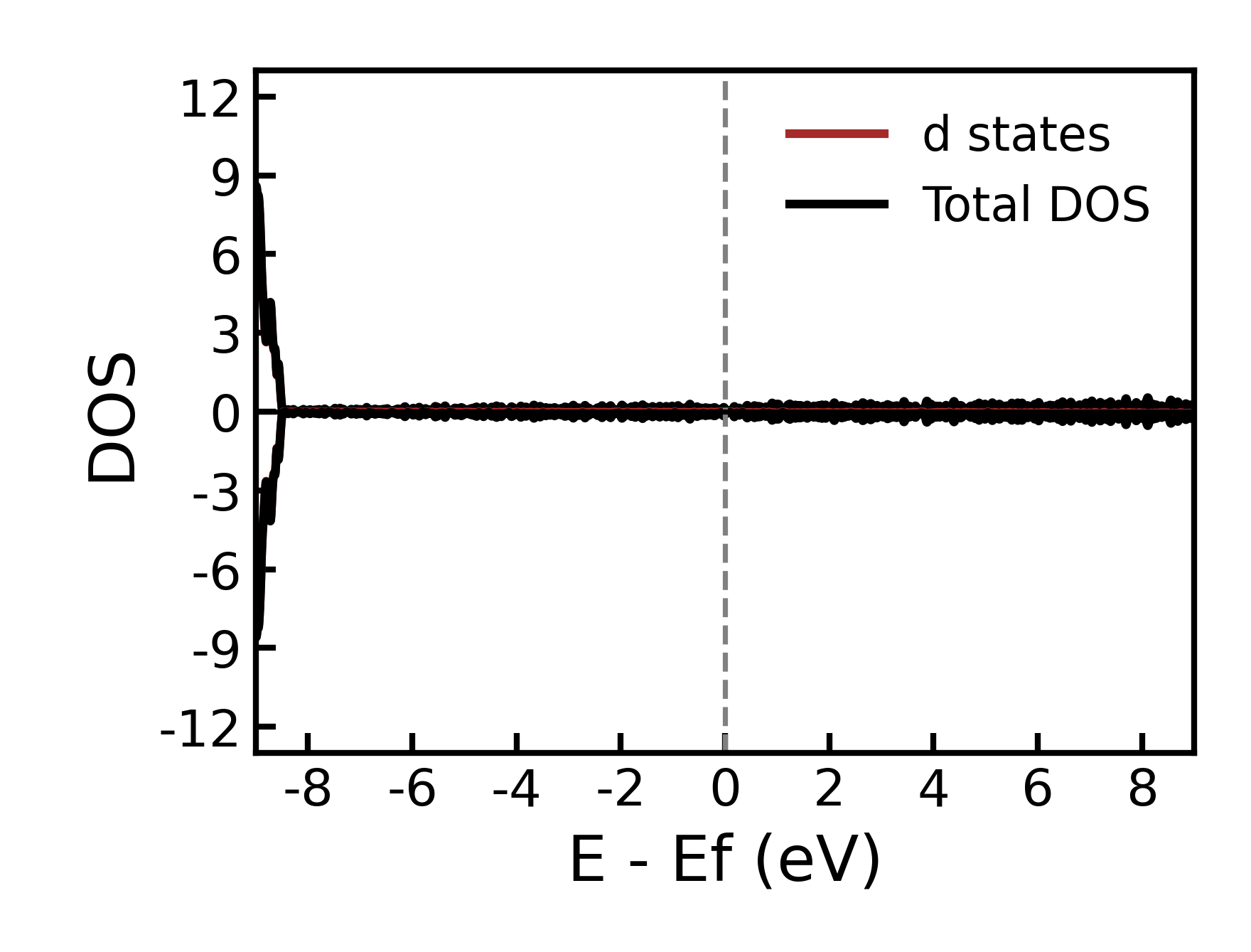}
    \caption{Cd DOS before and after nudging spin symmetry breaking.}
    \label{fig:dos_Gd}
\end{figure}

\FloatBarrier

\section{Numerical parameters for Density Functional Theory Calculations}
\label{sec_sup:numericals}

To study symmetry breaking and its relation to strong correlations we have used quantum mechanical methods based on the spin Density Functional Theory (spin-DFT) \cite{kohn1964,Kohn1965}. The $r^2\mathrm{SCAN}$ meta-GGA functional was used to recover exchange and correlation effects \cite{Furness2020}. Calculations were performed using the Projector-Augmented Wave (PAW) method \cite{paw}, implemented in the Vienna Ab-initio Simulation Package (VASP) software \cite{vasp1,vasp2}. Projectors were chosen to include all $f$ and $d$ electrons for every calculation within the PAW formalism, as detailed in table \color{blue}S1\color{black}.

It is well known that approximate density functionals suffer from self-interaction error (SIE), which is particularly significant for localized $d$ and, even more so, $f$ electrons. While the magnitude of SIE decreases when moving from LSDA to GGA and further to meta-GGA functionals such as $r^2\mathrm{SCAN}$, it is not fully removed at this level of theory. Importantly, recent works \cite{perdew-zunger-2025,Perdew2025,perdew2024} have demonstrated that reducing SIE often promotes spin symmetry breaking. Despite these limitations, density functional theory remains a standard and reliable framework for describing $d$- and $f$-electron systems, and has been extensively validated in the context of rare-earth materials and correlated quantum systems \cite{Zhang2022,Wang2025,Zanib2024,Giri2025}.

The plane-wave energy cutoff was set to $1.5$ times the maximum recommended cutoff energy (ENMAX) among the atomic potentials in the POTCAR file; for systems containing different atomic species, the largest ENMAX value was selected and multiplied by $1.5$ to define the cutoff. As mentioned in the main text, the most stable structural configurations were obtained by fully relaxing both the lattice parameters and the atomic positions for each system the nonmagnetic configuration. For structural relaxations, electronic convergence was achieved when the total energy difference between successive steps was below $1 \times 10^{-5}\ \text{eV}$, while ionic relaxation was considered converged when the forces on each atom were smaller than $0.01\ \text{eV}/\text{\AA}$.

In order to calculate $\Gamma$, as defined in equation \ref{eq:alpha_fermi_level}, we calculated the density of states within $r^2\mathrm{SCAN}$ and also used the uniform electron gas model. For the density of states (DOS) calculations, a stricter electronic convergence criterion was adopted, with the total energy difference set to $1 \times 10^{-6}\ \text{eV}$. In order to construct k-meshes to calculate $\Gamma$, we used the following k-meshes: $17 \times 17 \times 17$ for Ni and Cu, $18 \times 18 \times 18$ for Co; $22 \times 22 \times 22$ for Fe; $15 \times 15 \times 15$ for Pd and Ag; $19 \times 19 \times 10$ for Gd; $26 \times 26 \times 13$ for Zn; and $23 \times 23 \times 10$ for Cd. A $7 \times 7 \times 7$ mesh were used for both NiO and SmB$_6$. A $13 \times 13 \times 22$ k-mesh was used for VO$_2$ rutile while a $17 \times 10 \times 11$ was adopted monoclinic phase of VO$_2$. For EuB$6$ a $15 \times 15 \times 15$ k-mesh was used. These choices reflect the different lattice symmetries and ensure a consistent level of convergence across all systems.

Table \color{blue}S1 \color{black}shows the symmetry group of each material before and after relaxation, as well as the corresponding cell volumes. It also reports the number of atoms hosting correlated electrons in the unit cell and their electronic configurations, including the number of correlated electrons per atom for each material. In addition, the table presents the symmetry group obtained after full relaxation of the cell volume, cell shape, and atomic positions. Relaxations were done while maintaining the symmetric, non-magnetic, configuration. As discussed in the main text, to determine $N_e$ one multiplies the number of atoms with the number of correlated valence electrons for the specific materials. For Gd, for instance, we have $N_e = 2\: \mathrm{atoms} \times 7\: f\mathrm{\:electrons} = 14\: \mathrm{electrons}$, as the density of states near $\epsilon_F$ for Gd is mainly due to $f$ electrons.

\begin{table}[h]
\label{tab:supp_comp_details}
\caption{Columns represent: material studied, crystal symmetry before and after non-magnetic relaxation, cell volume $V_{\mathrm{cell}}$ after non-magnetic relaxation, number of atoms hosting correlated electrons $N_{\mathrm{atoms}}$ for each materials and valence electronic configurations used in the calculations for each material.}
\begin{ruledtabular}
\begin{tabular}{lccccc}
Material & Symmetry before & Symmetry after & $V_{\mathrm{cell}}$ (Å$^3$) & $N_{\mathrm{atoms}}$ & PAW valence configuration \\
\hline
Ni   & $Fm\bar{3}m$ & $Fm\bar{3}m$ & 41.409 & 4 & $3d^9 4s^1$ \\
Fe   & $Im\bar{3}m$ & $Im\bar{3}m$ & 20.300 & 2 & $3d^7 4s^1$ \\
NiO  & $Fm\bar{3}m$ & $Fm\bar{3}m$ & 556.875 & 32 & Ni: $3d^9 4s^1$, O: $2s^2 2p^4$ \\
VO$_2$ (r) & $P4_2/mnm$ & $P4_2/mnm$ & 57.778 & 2 & V: $3d^4 4s^1$, O: $2s^2 2p^4$ \\
VO$_2$ (m) & $P\bar{1}$ & $P\bar{1}$ & 131.782 & 4 & V: $3d^4 4s^1$, O: $2s^2 2p^4$ \\
Co   & $Fm\bar{3}m$ & $Fm\bar{3}m$ & 39.714 & 4 & $3d^8 4s^1$ \\
Pd   & $Fm\bar{3}m$ & $Fm\bar{3}m$ & 59.715 & 4 & $4d^{9} 5s^1$ \\
EuB$_6$ & $Pm\bar{3}m$ & $Pm\bar{3}m$ & 70.092
 & 1 & Eu: $5s^2 5p^6 4f^7 6s^2$, B: $2s^2 2p^1$ \\
SmB$_6$ & $Pm\bar{3}m$ & $Pm\bar{3}m$ & 557.684 & 8 & Sm: $5s^2 5p^6 4f^5 5d^1 6s^2$, B: $2s^2 2p^1$ \\
Gd   & $P6_3/mmc$ & $P6_3/mmc$ & 61.364 & 2 & $5s^2 5p^6 4f^7 5d^1 6s^2$ \\
Cu   & $Fm\bar{3}m$ & $Fm\bar{3}m$ & 45.618 & 4 & $3d^{10} 4s^1$ \\
Ag   & $Fm\bar{3}m$ & $Fm\bar{3}m$ & 68.999 & 4 & $4d^{10} 5s^1$ \\
Zn   & $P6_3/mmc$ & $P6_3/mmc$ & 28.570 & 2 & $3d^{10} 4s^2$ \\
Cd   & $P6_3/mmc$ & $P6_3/mmc$ & 44.443 & 2 & $4d^{10} 5s^2$ \\
\end{tabular}
\end{ruledtabular}
\end{table}

\section{Taking more occupied and unoccupied single-particle states into account}
\label{sec_sup:alphas}

In this section, we investigate the inclusion of states in the vicinity of the Fermi level, \textit{i.e.} occupied and unoccupied states within a certain energy range, to enhance the description of correlation encoded in $\Gamma$. We do this by integrating both density of states, from the density functional approximation and from the uniform electron gas model, using a gaussian filter. This naturally creates a free parameter that has to be adjusted from physical insight/intuition, as we shall see. We begin by defining $\Gamma_g$:

\begin{equation}
    \label{eq:alpha_2}
    \Gamma_g = \sqrt{ \left(\int_{-\infty}^{0} \mathrm{D}(\epsilon) e^{-\left(\frac{\epsilon}{\delta}\right)^2} d\epsilon \right) \left( \int_{0}^{\infty} \mathrm{D}(\epsilon) e^{-\left(\frac{\epsilon}{\delta}\right)^2} d\epsilon \right) \left( \int_{-\infty}^{0} \mathrm{D}_{\mathrm{unif}}(\epsilon) e^{-\left(\frac{\epsilon}{\delta}\right)^2} d\epsilon \right)^{-1} \left( \int_{_0}^{\infty} \mathrm{D}_{\mathrm{unif}}(\epsilon) e^{-\left(\frac{\epsilon}{\delta}\right)^2} d\epsilon \right)^{-1}}.
\end{equation}

This expression is designed to be properly small when there are few occupied states, or few unoccupied states, within energy range $\delta$ of the Fermi level. As one can see, the new parameter $\delta$ has the units of energy and controls the width of the gaussian filter, or in other words, the energy range in which states are being integrated. As $f$ electrons are more localized in space and energy if compared to $d$ electrons, we made a systematic study using different values of $\delta$ for $f$ ($\delta_f$) and $d$ ($\delta_d$) electron materials. After that, we calculated the pearson correlation between the $\Gamma_g$ values obtained and the respective $\Delta E$ for the specific material. The results are shown below:

\begin{figure}[h]
    \centering
    \includegraphics[width=0.8\linewidth]{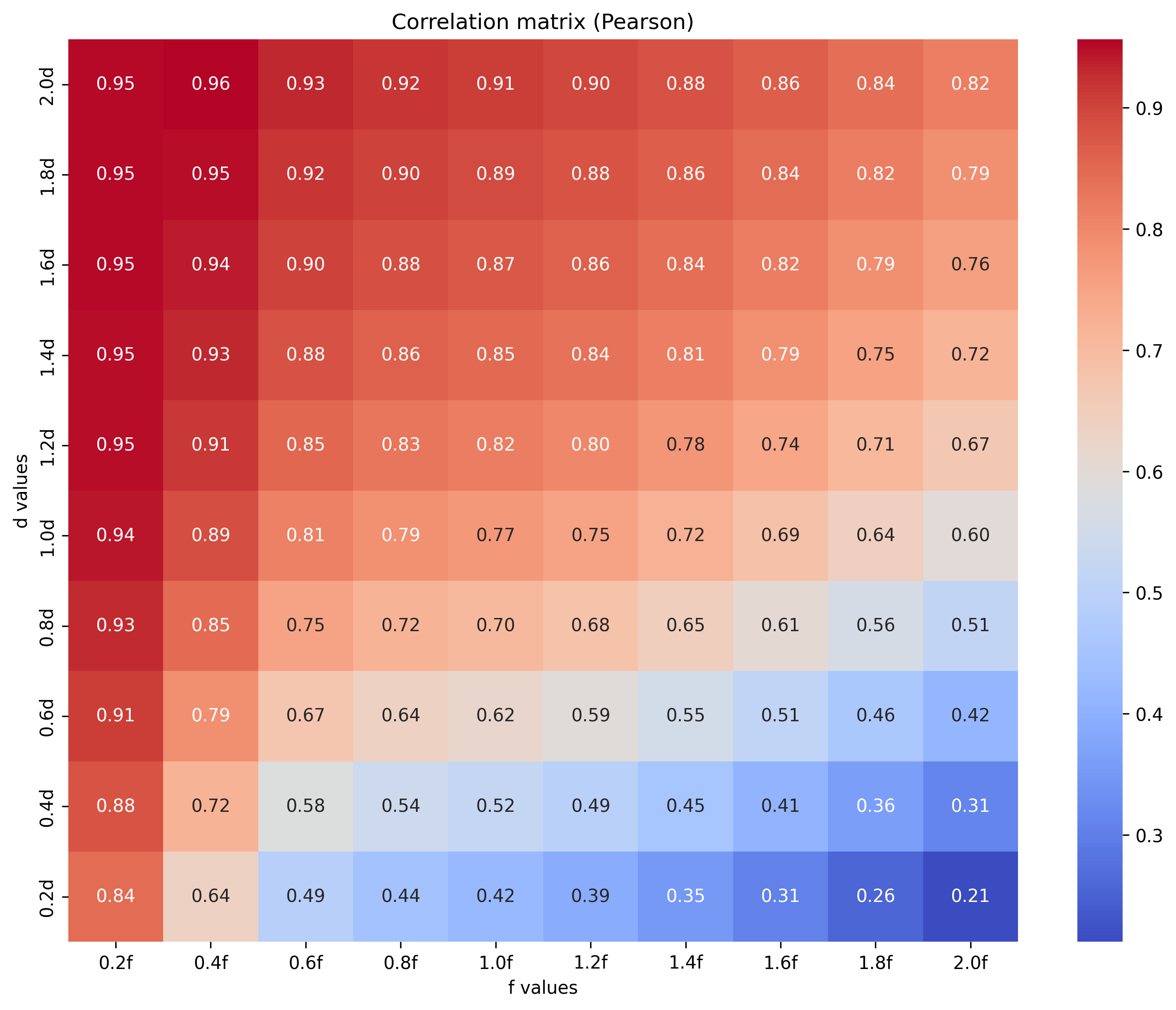}
    \caption{Pearson correlation matrix between $\Gamma_g$ and $\Delta E$ for different values of $\Gamma$ for \textit{d} and \textit{f} electron materials separately.}
    \label{fig:correlation_matrix}
\end{figure}

By analysing figure \ref{fig:correlation_matrix}, one can see that, for the same $\delta$ for \textit{d} and \textit{f} electrons (main diagonal) the best Pearson correlation coefficient reached is 0.84. Nevertheless, by decreasing $\delta$ for \textit{f} electrons, which reflects the more local nature of \textit{f} electrons in both real space and energy, Pearson correlation is able to reach a remarkable value of 0.96. This indicates that, indeed, $\Gamma_g$ (when calculated with the proper $\delta$ value -- a good number of occupied and unoccupied states) and $\Delta E$ are measuring the same physical quantity: the amount of strong correlation that is not well described by a density functional approximation in the symmetric state.

We show below a $\Gamma_g$ vs $\Delta E$ plot for $\delta_d = 2.0\: eV$ and $\delta_f = 0.4\: eV$ to illustrate a $\delta$ configuration of $\Gamma_g$ that yields a great Pearson correlation coefficient (0.96), followed by a $\Gamma_g$ vs $\Delta E$ plot for $\delta_d = 0.4\: eV$ and $\delta_f = 2.0\: eV$ to illustrate the case with a bad Pearson correlation (0.31):
\begin{figure}[h]
    \centering
    \includegraphics[width=0.65\linewidth]{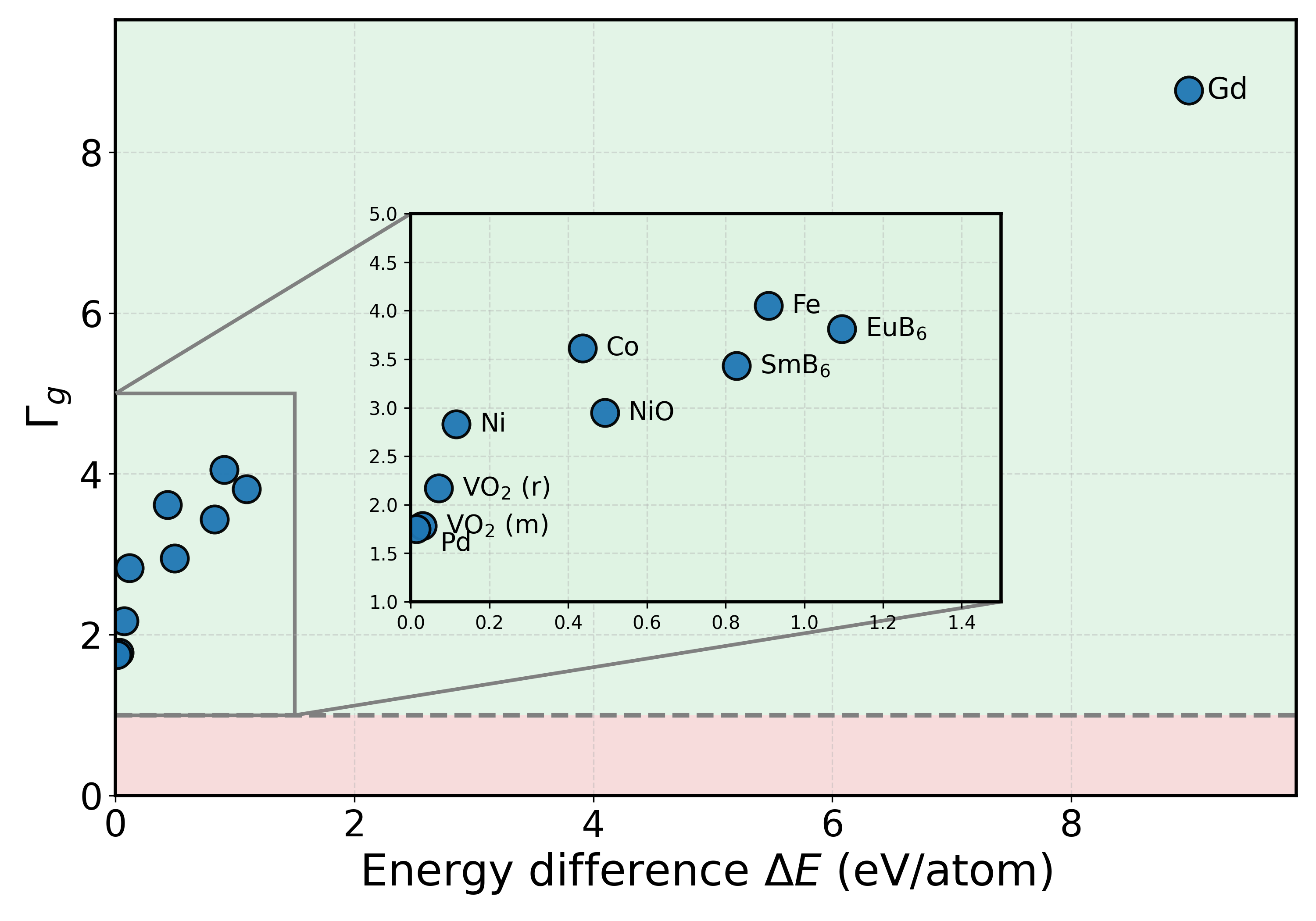}
    \caption{$\Gamma_g$ \textit{vs} $\Delta E$ plot. Green (red) region is for strongly (normally) correlated system. This plot employs $\delta_d = 2.0\:eV$ and $\delta_f = 0.4\:eV$. Here the Pearson correlation coefficient is 0.96. When excluding Gd from the analysis, the Pearson correlation coefficient remains high (0.89), indicating a strong statistical correlation.}
    \label{fig:alpha_gau_bom}
\end{figure}

\begin{figure}[h]
    \centering
    \includegraphics[width=0.65\linewidth]{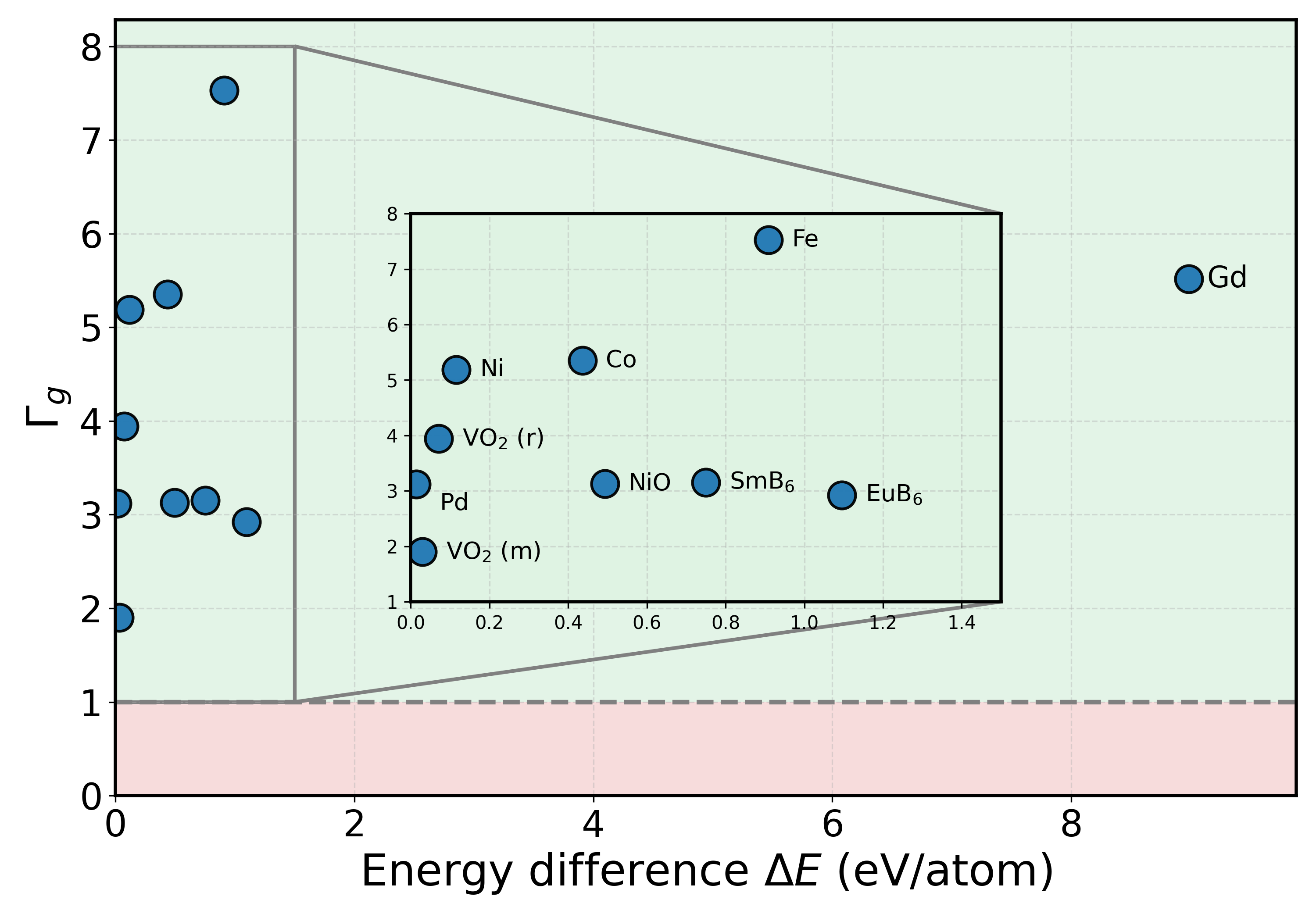}
    \caption{$\Gamma_g$ \textit{vs} $\Delta E$ plot. Green (red) region is for strongly (normally) correlated system. This plot employs $\delta_d = 0.4\:eV$ and $\delta_f = 2.0\:eV$. Here the Pearson correlation coefficient is 0.31.}
    \label{fig:alpha_gau_ruim}
\end{figure}

Below, we show the actual $\Gamma_g$ for each case shown in figures \ref{fig:alpha_gau_bom} and \ref{fig:alpha_gau_ruim}:
\pagebreak
\begin{table}[h]
\centering
\caption{Values of $\Gamma_g$ and $\Delta E$ for two different parameter regimes.}
\begin{tabular}{lccc}
\hline
\hline
Material &
$\Gamma_g$ (2.0d–0.4f) &
$\Gamma_g$ (0.4d–2.0f) &
$\Delta E$ (eV/atom) \\
\hline
Ni   & 2.829 & 5.189 & 0.116 \\
Fe   & 4.050 & 7.530 & 0.909 \\
NiO  & 2.948 & 3.127 & 0.493 \\
VO$_2$ (r) & 2.171 & 3.943 & 0.071 \\
VO$_2$ (m) & 1.780 & 1.901 & 0.030 \\
Co   & 3.614 & 5.351 & 0.436 \\
Pd   & 1.750 & 3.121 & 0.014 \\
EuB$_6$ & 3.812 & 2.921 & 1.095 \\
SmB$_6$ & 3.433 & 3.154 & 0.750 \\
Gd   & 8.768 & 5.514 & 8.985 \\
\hline
Cu   & 1.151 & 0.620 & 0.000 \\
Ag   & 0.465 & 0.424 & 0.000 \\
Zn   & 0.486 & 0.375 & 0.000 \\
Cd   & 0.456 & 0.326 & 0.000 \\
\hline
\hline
\end{tabular}
\end{table}

The value of $\Gamma_g$ for Cu is slightly larger than 1 for $\delta_d = 2.0\: eV$ and $\delta_f = 0.4\: eV$, although it remains very close to this limit. This happens because of the relatively high density of states around $-1.8\: eV$, as shown in figure \ref{fig:Cu_dos}. Despite the imbalance between occupied and unoccupied states, a large value of $\delta$ leads to a significant weight from the Gaussian factor in equation \ref{eq:alpha_2}, such that even the low density of unoccupied states contributes non-negligibly. As a result, when this contribution is multiplied by the integral over the occupied states (both evaluated with $r^2\mathrm{SCAN}$), the numerator of Eq.~\ref{eq:alpha_2} becomes slightly larger than the denominator, yielding $\Gamma_g > 1$. Nevertheless, the correlation is still classified as normal, since $\Gamma_g$ is close to 1, the large portion of the density of states is far from $\epsilon_F$ -- approximately $1.8 \: eV$ -- and the system does not undergo symmetry breaking, i.e., $\Delta E = 0$.

\section{Comparisson with the Stoner Criteria}
\label{sec_sup:stoner}

Symmetry instabilities and itinerant ferromagnetic phase transitions are well described by the Stoner criterion \cite{mohn2002magnetism,tong_statphys}. Within this framework, the nonmagnetic (paramagnetic) state corresponds to $\langle \mathbf{S}_i \rangle = 0$, while a ferromagnetic state is characterized by a finite local moment, $\langle \mathbf{S}_i \rangle \neq 0$. At zero temperature, the instability of the symmetric phase is governed by the dimensionless parameter $I \mathrm{D}(\epsilon_F)$, where $I$ is the Stoner exchange parameter and $\mathrm{D}(\epsilon_F)$ is the density of states at the Fermi level. When $I \mathrm{D}(\epsilon_F) > 1$, the paramagnetic state becomes unstable, signaling spontaneous symmetry breaking of the SU(2) spin-rotation symmetry.

From a density functional theory perspective, this instability can be understood more precisely through the spin susceptibility, which can take the Stoner form $\chi = \chi_s/[1 - I \mathrm{D}(\epsilon_F)]$, where $\chi_s > 0$ is the susceptibility of the Kohn–Sham non-interacting system \cite{MacDonald1976}. This result provides a practical route: one can predict the tendency toward ferromagnetism by analyzing the stability of the nonmagnetic Kohn–Sham state within standard (normally correlated) DFT. Indeed, for transition metals, this approach correctly identifies Fe, Co, and Ni as systems with $I \mathrm{D}(\epsilon_F) \gtrsim 1$, while for others it remains below unity \cite{moruzzi1978calculated,martin2004electronic}. For Pd, particularly, the product remains close to 1 (0.78), indicating proximity to instability, as already discussed \cite{moruzzi1978calculated,martin2004electronic}.

However, this predictive success should be interpreted with care. The instability of the nonmagnetic Kohn–Sham state does not imply that the corresponding symmetric physical state is itself normally correlated. Rather, the symmetric state can be viewed as energetically stabilized by strong correlation effects, even if it is not experimentally realized due to slow magnetic fluctuations. Calculations with normally-correlated  density functional approximations show that normal correlation cannot always stabilize the symmetric or non-magnetric state.


From the perspective of a strongly correlated system described by the Hubbard model, the usual interpretation of the Stoner instability is formulated in terms of the on-site interaction $U$. In this context, the emergence of magnetism is associated with the existence of a critical interaction strength $U_c$, above which the paramagnetic phase becomes unstable toward magnetic ordering. Within this view, strong electronic correlations are directly encoded in the magnitude of the local Coulomb repulsion $U$. In contrast, in our perspective the Stoner criterion is discussed by emphasizing the role of the density of states at the Fermi level. Here, the signature of strong correlations is not attributed directly to a large interaction parameter, but rather to an enhanced $\mathrm{D}(\epsilon_F)$, which itself leads to stronger electronic interactions. In this sense, a high density of states at the Fermi level enhances the tendency toward symmetry breaking.

Thus, our results can be reinterpreted within the framework of the Stoner criterion from a different perspective on the role of correlation, emphasizing near-degeneracies and enhanced density of states at the Fermi level in the KS system as effective signatures of strong correlation in the interacting system, and providing a natural connection between strong correlation and symmetry breaking. It is interesting to notice that both frameworks give hints about the symmetry broken state only by calculating the symmetric state.

\putbib
\end{bibunit}
\end{document}